# WAKE-NET: A 3D-Wake-Aware Economic Turbine Layout and Cabling Optimization Framework for Multi-Capacity Multi-Hub-Height Wind Farms Serving Grid-Scale and Industrial Power Systems

Ann Mary Toms[a], Xingpeng Li[a]

[a]*University of Houston, Department of Electrical and Computer Engineering, Houston, TX, 77204-4005*

---

**Abstract**

The global transition towards renewable energy has accelerated the deployment of utility-scale wind farms, increasing the need for accurate performance and economic assessments. Although wind energy offers substantial potential for carbon emission reduction, investment decisions are highly sensitive to predicted annual energy production and economic profitability. Conventionally wind farm analyses often estimate turbine power output based solely on incoming wind conditions, neglecting wake interactions between turbines. These wake effects can significantly reduce downstream turbine performance, leading to overestimation of energy yield and financial returns. This study proposes WAKE-NET, a 3D wake-aware optimization framework that integrates turbine layout optimization, turbine capacity selection, cable routing, and hub height diversification within a unified profit-driven formulation. Unlike traditional approaches that assume a uniform hub height and turbine capacities or ignore wake dynamics, the proposed framework accounts for wake-induced power losses during optimization. A benchmark wake-ignorant model is also evaluated to quantify the impact of neglecting wake interactions. Results indicate that the wake-ignorant optimization can significantly overestimate annual profits, while the use of multiple hub heights and capacities reduce wake overlap and improve spatial utilization. Overall, the findings demonstrate that wake-aware optimization coupled with hub height and capacity diversification provides more reliable energy yield prediction and economic assessment, offering valuable guidance for large-scale wind farm planning and investment.

---

Nomenclature

| | | | |
|---|---|---|---|
| Abbreviations | | | |
| AEB | Annual Economic Benefit | AEP | Annual Energy Production |
| APB | Annual Production Benefit | MST | Minimum Spanning Tree |
| | | | |
| Indices and Sets | | | |
| $i$ or $j$ | Turbine number indices | $T$ | Number of turbines |
| $U$ | Set of all wind speed bins | $\Theta$ | Set of all wind direction bins |
| Variables | | | |
| $A_{overlap}$ | Overlap area between the expanding wake cross section and the downstream rotor disk | $d$ | Three-dimensional distance between wake centerline and downstream rotor center |
| $f_{ij}$ | Wake overlap fraction | $P_{w_i}$ | Electrical power output of turbine $i$ |
| $v_{h_1}$ | Velocity of wind measured by the anemometer at height $h_1$ | $v_{h_2}$ | Velocity of wind at hub height $h_2$ |
| $v_{i,h_2}$ | Velocity of wind at turbine $i$ at hub height $h_2$ | $v_{j,eff}$ | Effective wind speed at turbine $j$ after wake interactions |
| $\delta_{ij}$ | Wake-induced velocity deficit contribution | $\Delta_c$ | Wind speed deficit along the wake centerline |
| $\Delta_j$ | Combined wake deficit acting on turbine $j$ | | |
| Parameters | | | |
| $A_w$ | Swept area of turbine rotor | $C_p$ | Power coefficient |
| $C_T$ | Thrust coefficient | $D_i$ | Rotor diameter of turbine $i$ |
| $h_1$ | Height at which the anemometer is installed | $h_2$ | Hub height of the wind turbine |
| $I$ | Turbulence intensity | $k$ | Wake expansion coefficient |
| $p_u$ | Probability of wind speed $u$ | $p_\theta$ | Probability of wind coming from direction $\theta$ |
| $P_{t_i}^{rated}$ | Rated Power capacity of turbine $i$ | $R_i$ | Rotor radius |
| $R_{wake}$ | Wake radius | $x_{ij}$ | Downstream distance between turbines $i$ and $j$ |
| $y_{ij}$ | Horizontal/ lateral distance between turbines $i$ and $j$ | $z_0$ | Roughness length |
| $z_i$ | Hub height of turbine $i$ | $\alpha$ | Axial induction factor |
| $\theta$ | Wind direction | $\rho_w$ | Air density at the hub height |
| Economic Parameters | | | |
| $C_{cable}$ | Overall cable infrastructure investment | $C_{elec}$ | Cost of electricity ($/kWh) |
| $C_{land}$ | Overall land investment | $C_{turbine}$ | Overall turbine investment |
| $c_{capex,t_i}$ | Capital cost of turbine $j$ | $c_{land}$ | Land cost ($/sq.m) |
| $c_{O\&M}$ | Annual turbine operations and maintenance cost | $L_{MST}$ | Total length of cable |
| $N_{days}$ | Number of days in a year | $N_{hours}$ | Number of hours in a day |
| $N_{year}$ | Wind farm lifetime | $r$ | Interest rate |
| $(x_{min}, x_{max})$, $(y_{min}, y_{max})$ | Wind farm spatial boundary | | |

## 1. Introduction

Wind energy has become a cornerstone of the global transition towards sustainable power systems. As nations strive to meet net zero targets and reduce their reliance on fossil fuels, wind energy deployment has expanded rapidly. Both onshore and offshore wind farm installations now play a critical role in meeting growing electricity demands while curbing greenhouse emissions.

Despite its advantages, wind energy production is significantly affected by wake interactions and suboptimal turbine layouts. Wake-induced velocity deficits and increased turbulence can reduce the power output of downstream turbines, leading to energy loss at the wind farm level. Wake-related losses typically range from 4-25% of total power production with larger offshore wind farms experiencing an average loss on the order of 10-20% [1]. Onshore studies in North Texas report lower losses of approximately 2-4% [2], while offshore developments like the Danish Energy Island indicate losses of 8.6-10.1% [3]. In extreme cases involving poor turbine spacing, total power losses may reach up to 25% [4], with individual downstream turbines experiencing power reductions of up to 46% due to wake effect [5].

These significant losses underscore the importance of wind farm optimization, which remains a challenging problem due to complex wake interactions, nonlinear turbine behavior, and competing economic and spatial constraints. Effective optimization strategies are therefore essential to accurately predict energy yield and ensure the long-term viability of large-scale wind energy projects.

Wake losses in wind farms are primarily caused by the aerodynamic interactions between wind turbines wherein the extraction of kinetic energy by an upstream turbine generates a region of reduced wind speed and increased turbulence downstream. This wake can persist for several rotor diameters, significantly degrading the performance of downstream turbines operating within its influence. The magnitude and recovery of wake effects are governed by multiple factors, including ambient wind speeds, wind direction, turbulence intensity, and turbine spacing and layout [6]. As a result, wake-induced power losses are highly site-specific and strongly dependent on the wind farm configuration.

To quantify these effects, a wide range of wake models with varying levels of fidelity have been developed. Empirical models such as the Jensen model [7] and the Ainslie eddy-viscosity model [8], are widely used due to their computational efficiency. However, these simplified models rely on empirical assumptions that may limit accuracy under complex atmospheric conditions or closely spaced turbine arrangements. Higher fidelity approaches, including large eddy simulations [9], and more recently machine learning-based wake models[10], offer improved predictive capability but at a significantly higher computational cost [11].

In poorly optimized wind farms, inadequate turbine layout can exacerbate wake interactions, leading to severe performance penalties. Such configurations may result in cumulative energy losses across the farm, reduced capacity factor, and diminished economic returns, underscoring the necessity of incorporating wake-aware modeling and optimization in wind farm design.

While previous studies have explored the use of multiple hub heights of turbines for wake mitigation and cost-of-energy reduction, the integration of hub height and capacity diversification within a unified wake-aware economic optimization framework remains limited.

### *1.1. Literature Review*

The wake effect in wind farms is primarily caused by the aerodynamic interactions between wind turbines, where the extraction of kinetic energy by an upstream turbine generates a region of reduced wind speeds and enhanced turbulence downstream[12]. This disturbed flow region, known as wake, can extend for several rotor diameters and directly impinge on downstream turbines operating within their path. The resulting velocity deficit reduces the available wind energy at downstream turbines [6], while increased turbulence intensity [13] leads to higher unsteady loading and accelerated structural fatigue [14]. Recent studies have explored wake-aware layout optimization aimed at minimizing turbulence induced fatigue loading and improved lifetime [15]. Consequently, wake interactions not only reduce power production but also affect turbine reliability and operational lifespan. Field measurements from large offshore wind farms such as Horns Rev [16] consistently demonstrate sustained wake effects across multiple turbine rows. Although wake induced turbulence can significantly influence structural fatigue loading, and turbine lifespan, the present study focuses primarily on wake induced velocity deficits and their impact on energy production and economic performance. Detailed fatigue load modeling and turbulence transport analysis are beyond the scope of the current work.

Atmospheric conditions play a critical role in wake behavior, with atmospheric stability and wind turbulence intensity dramatically influencing wake dissipation. Convection conditions enhance turbulent mixing and accelerate wake recovery, whereas stable atmospheric conditions suppress mixing and prolong wake persistence. Experimental and numerical studies show that wake growth rate under convective conditions can be 2.4 times larger than those observed under stable conditions [17], with power deficits reaching up to 50% during stable periods [6]. This strong dependence on environmental conditions underscores the necessity of accurate wake modeling for reliable wind farm performance assessment.

Wind farm layout optimization relies heavily on wake models to estimate turbine interactions and power losses, making wake modeling a central component of wind farm design studies [18]. The most commonly used engineering wake models for wind farm studies are the Jensen, Larsen, Frandsen, and Dynamic Wake Meandering models which are widely adopted due to their computational efficiency. These models rely on simplifying assumptions such as overestimation of flow characteristics [19], uniform or gaussian wake profiles [20], which can limit accuracy under complex operating conditions. Despite these limitations, analytical and semi-empirical wake models provide a balance between model fidelity and computational simplicity, making them well suited for layout optimization and large-scale wind farm studies [21].

Higher fidelity approaches, particularly Large Eddy Simulation (LES), offer detailed resolution of wake turbulence and recovery processes, significantly improving predictive accuracy [22]. However, the high computational expense of LES restricts its use in large-scale optimization, prompting the development of multi-fidelity and hybrid optimization frameworks [23]. More recently, data-driven and machine learning-based wake models have emerged, achieving substantial improvements in wake and power prediction accuracy by learning complex flow interactions from high-fidelity simulation data [10]. These approaches, however, require large training datasets and face challenges related to generalization across varying atmospheric and layout conditions.

The primary objective of wind farm layout optimization has evolved from simple energy maximization to a broader goal of minimizing costs while maximizing energy production. Early studies focused primarily on maximizing power output through optimal turbine placement [24], but subsequent research demonstrated that energy yield alone does not ensure economic viability [25]. Modern optimization frameworks therefore adopt a multi-objective formulation that balances electricity generation costs, wake-induced losses and turbine configurations [26]. These optimization strategies reduce wake interactions by strategically positioning turbines to minimize downstream velocity deficits [5] and by incorporating wake-aware models in the design process [27].

A wide range of optimization techniques have been extensively applied to this problem including genetic algorithms, particle swarm optimization, gradient-based, and hybrid methods [28]. Practical layout optimization also incorporates constraints related to land use [29], structural loading [30], economic constraints and technological constraints [31]. These constraints can significantly limit achievable energy gains, underscoring the complexity of wind farm layout optimization in real-world applications.

Wind farm optimization studies incorporate economic performance metrics along with energy production to better reflect real-world investment objectives. Rather than maximizing energy yield alone, modern frameworks evaluate metrics like levelized cost of energy, net present value, internal rate of return, and profitability index, enabling wind farm designs that align with market-driven profitability [32]. Optimization models typically account for capital cost, operations, and maintenance costs and energy yield, with costs strongly influenced by turbine size, site selection, water depth, distance to shore and electrical infrastructure [33]. Neglecting wake effects has been shown to result in significant profit overestimation including multi-million dollar losses over project lifetimes [34]. Simplified wake assumptions further undermine the reliability of techno economic assessments, as different wake models can produce different estimates of wind farm generation potentially differing by up to 70% in its capacity factor [35]. These uncertainties are particularly critical for large-scale and offshore wind projects where high capital investments amplify the financial consequences of inaccurate performance prediction.

Most wind farm studies traditionally assume uniform turbine configurations with uniform hub heights and capacities primarily due to its modeling simplicity and ease of implementation. However, recent research challenges this assumption, demonstrating that wind farms with non-uniform turbine designs have improved energy yield and economic performance [36]. Researchers have developed a 3D greedy optimization approach using turbines at multiple hub heights and demonstrated that vertical staggering turbines could increase total power production while reducing cost per unit power output, particularly under complex terrain conditions [37]. Nevertheless, hub height strongly influences wind speed exposure, vertical sheer and wake overlap as taller turbines experience higher mean wind speeds [38]. Introducing hub height variations can reduce wake interference by vertically staggering turbines, thereby limiting wake overlap and exploiting different wind profiles. Studies show that such configurations can increase downstream turbine power output and reduce wake-induced losses, with reported improvements in energy production and cost reduction exceeding 10% in favorable conditions [39]. Researchers have also shown that combining different hub heights with wake-aware control strategies improved wind farm efficiency and reduced cost of energy in real wind farm case studies [40].

Research in wind farm modeling and optimization has advanced significantly; however, several gaps remain in existing literature. Firstly, most optimization studies rely on simplified wake models. While computationally efficient, these models struggle to accurately capture complex wake interactions in large or densely packed wind farms. Although high-fidelity approaches like LES offer improved accuracy, their high computational time hinders integration into large, iterative optimization frameworks. Secondly, many studies treat wake-aware energy modelling and economic performance assessment as two separate problems, resulting in limited insight as to how wake-induced losses affect long-term profitability and investment. Thirdly, profit-driven optimization frameworks typically assume homogeneous turbine configurations with uniform hub heights despite growing evidence that hub height diversification can mitigate wake interference and improve performance. Existing studies tend to optimize turbine layouts, turbine capacities, or turbine hub heights in isolation, neglecting the strong coupling between these variables. As a result, current approaches fall short of providing an integrated, robust yet economical solution for wind farm optimization.

### *1.2. Objectives and Contributions*

The primary goal of this work is to provide a scalable, wake-aware optimization framework that supports informed decision-making in modern utility-scale wind farm planning. The main contributions include the development of an integrated optimization framework that combines wake-aware layout design, hub height diversification, turbine capacity selection and cable routing within a profit-driven objective function.

This paper proposes a wake-aware network optimization framework called WAKE-NET, which integrates aerodynamic, spatial, and economic considerations within a unified framework. WAKE-NET explicitly accounts for wake-induced power losses while jointly optimizing wind farm layout and turbine characteristics. The framework is designed to determine the optimal land area required for a wind farm, turbine placement within that area, turbine capacity selection, and electrical cable routing, all under the objective of maximizing long-term profitability. By embedding the wake interactions directly into the optimization process, WAKE-NET improves the accuracy of energy yield estimation and mitigates profit overestimation associated with simplified wake assumptions.

A key novelty of the proposed framework is the incorporation of turbine heterogeneity with multiple hub heights enabling vertical staggering of turbines to reduce wake overlap. WAKE-NET evaluates hub height diversification alongside layout optimization rather than as a post-processing variable, allowing the coupled effect of wake dynamics and turbine configuration to be captured realistically. The framework further leverages historic, site-specific wind data from both onshore and offshore locations. Collectively, these features enable a more reliable assessment of both aerodynamic performance and economic profit, particularly for large wind farms where wake losses and capital investment risks are substantial.

Unlike prior studies that primarily focused on minimizing cost of energy or improving wake mitigation through hub height variation alone, the proposed WAKE-NET framework integrates wake-aware economic optimization, turbine capacity selection, land utilization and cable routing within a unified profit driven formulation.

## 2. Methodology

The WAKE-NET framework is developed to maximize the annual economic benefit (AEB) of a utility-scale wind farm that can be connected to the bulk grid or a large energy-intensive hyperscale microgrids like that used for datacenters. It jointly optimizes layout, turbine selection and electrical infrastructure while explicitly accounting for wake interactions. Power production is estimated using the Jensen wake model with squared wake superposition. The effect of wake on the power production is evaluated concurrently during layout optimization to avoid overestimation of revenue. The framework uses five years' worth of site-specific hourly historical wind speed and direction data from different onshore and offshore locations across the United States. Optimization variables include turbine positions, turbine capacities, hub height, substation location, wind farm area, and cable routing. The cable routing is optimized using a minimum spanning tree approach. A genetic algorithm is employed to find the best solution, assuming a fixed number of turbines, constant electricity price, and full grid absorption of the generated power.

## 2.1. Wind Resource Characterization

Wind condition across the U.S. varies greatly due to the influence of large-scale weather systems including cold and warm fronts. As a result, understanding site-specific wind characteristics is a critical first step in selecting suitable locations for wind farm development. Accurate knowledge of wind speeds and directions enable more efficient turbine configurations and provide insight into wake interactions between turbines, which directly affect overall farm performance.
To capture realistic wind behavior, this research uses publicly available datasets maintained by the National Centers for Environmental Information (NCEI) [41] and the National Data Buoy Center (NDBC) [42]. These datasets were selected because they provide extensive historical hourly wind speed and directional measurements with minimal missing data. Fig.1 shows the geographic locations and station IDs corresponding to the selected onshore and offshore sites.

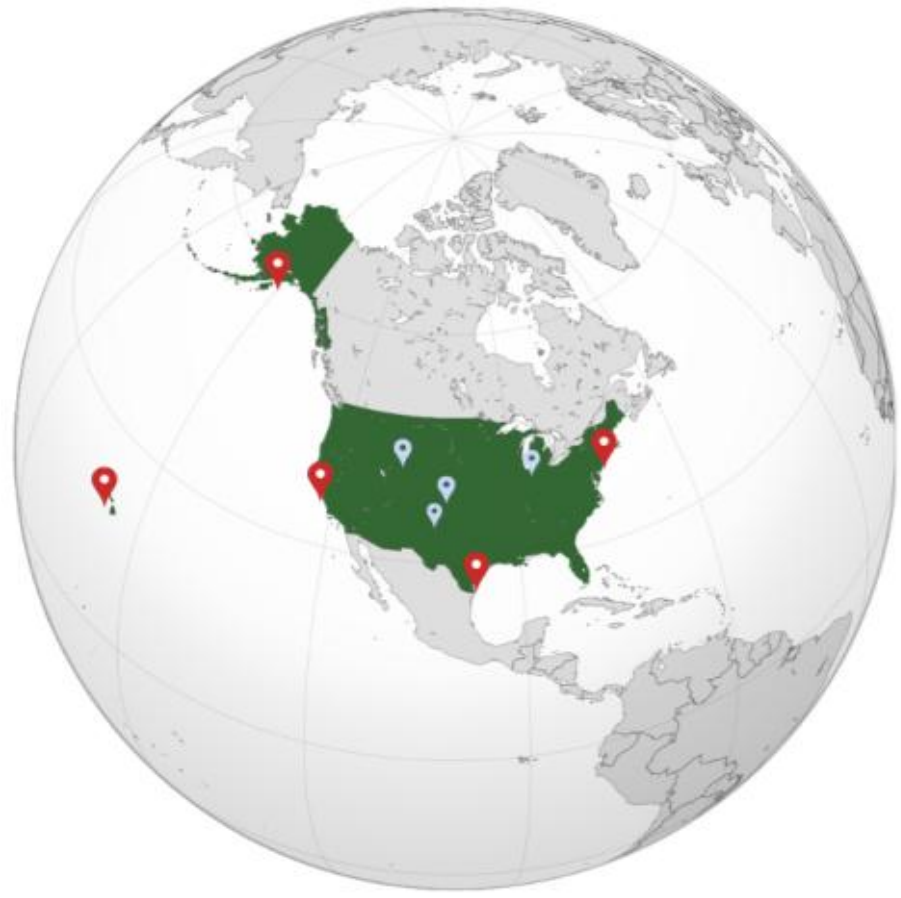

| *Region* | *NDBC Station ID for offshore* | *NCEI Station ID for onshore* |
|---|---|---|
| ***AK*** | 46001 (56°17'46" N 148°1'37" W) | - |
| ***HI*** | 51000 (23°32'3" N 153°45'9" W) | - |
| ***TX*** | 42020 (26°58'11" N 96°40'45" W) | USW00023047 (35°13′10″N 101°42′21″W) |
| ***CA*** | 46011 (34°56'14" N 120°59'58" W) | - |
| ***NJ*** | 44091 (39°46'19" N 73°46'8" W) | - |
| ***KS*** | - | USW00013985 ( 37°45′47″N 099°57′56″W) |
| ***WY*** | - | USW00024057 ( 41°48′12″ N, 107°12′0″ W) |
| ***IN*** | - | USW00014835 (35°13′10″N 86°56′13″ W) |

Fig. 1. Geographical map along with station IDs of the five offshore: Alaska, Hawaii, Texas, California, New Jersey, and four onshore: Texas, Kansas, Wyoming, Indiana regions analyzed in this study.

Although wind farm siting is influenced by multiple factors such as environmental constraints and land use restrictions, this work focuses exclusively on regions with high wind potential as identified using wind resource maps from the National Renewable energy Laboratory (NREL) [43]. Five years' worth of hourly historical data are used to ensure robust wind characterization and to reduce bias associated with anomalous years, such as those affected by extreme weather events.

Offshore wind speeds are measured near sea level, while onshore measurements are typically taken at approximately 10m height. Since most wind turbines operate at hub heights of 80m or higher [44], wind speeds are extrapolated to turbine hub heights using the logarithmic wind profile given by (1) [45]:

$$v_{h_2} = v_{h_1} \frac{ln\left(\frac{h_2}{z_0}\right)}{ln\left(\frac{h_1}{z_0}\right)} \tag{1}$$

where $h_2$ is the hub height of the wind turbine; $h_1$ is the height at which the anemometer is installed; $v_{h_1}$ is velocity of wind measured by the anemometer; $z_0$ is the roughness length of the sea which is taken as 0.0002m for open sea and 0.03m for open land.

### 2.2. Electrical Power Generation from Wind Turbines

Accurately modelling the power that can be generated from a wind turbine is essential for estimating how much electricity can realistically be generated at a given site and for selecting turbine capacities that match the local wind resource. Oversized turbines installed in regions with low wind may operate far below their rated capacity, leading to inefficient investment.

The power available in the wind is calculated using (2):

$$P_w = \frac{1}{2} \times \rho_w \times A_w \times \left(v_{h_2}\right)^3 \times C_p \tag{2}$$

where $\rho_w$ is density of air at the hub height; $A_w$ swept area of the turbine rotor; $v_{h_2}$ is velocity of wind at the hub height; $C_p$ is the coefficient of power [46].

The actual electrical power output is then determined using the turbine power curve which accounts for the cut-in, rated, and cutout speeds as shown in (4):

$$P_{w_i} = \begin{cases} 0 & , v_{h_2} \leq v_{cutin} \; or \; v_{h_2} \geq v_{cutout} \\ \min\left(P_w\left(v_{i,h_2}\right), P_{t_i}^{rated}\right) & , otherwise \end{cases} \tag{4}$$

Air density plays a key role and differs between onshore and offshore sites due to variation in altitude and temperature, with offshore locations typically experiencing denser air [47]. In this study, turbine hub heights are assigned based on turbine capacity, with larger turbines installed at higher elevations. All turbines are assumed to actively yaw and align with incoming wind direction.

## 3. Modelling Wake Effect

Wind farm layout optimization requires repeated estimation of turbine power for different layout configurations. This process involves recalculating the effective wind speed at each turbine while accounting for wake interactions from upstream turbines. To model these effects, this section first introduces the Jensen wake model which estimates the wind speed deficit in the wake region generated by a single turbine. The formulation is then extended to incorporate wake superposition, allowing the combined influence of multiple upstream turbine wakes on downstream turbines to be quantified. This approach enables efficient evaluation of wake-induced power losses during layout optimization.

### 3.1. Jensen Wake Model for a Single Turbine

The Jensen wake model is one of the earliest and most widely adopted analytical wake models due to its simplicity and computational efficiency. Consider an upstream turbine $i$ and downstream turbine $j$ as illustrated in Fig.2. The turbine positions are defined in a wind-aligned coordinate system for a given wind direction $\theta$, where $x_{ij}$ denotes the downstream distance between the turbines and $y_{ij}$ represents the lateral offset. When $x_{ij} \leq 0$, turbine $j$ is not located in the wake of turbine $i$, and therefore no wake interaction is assumed.

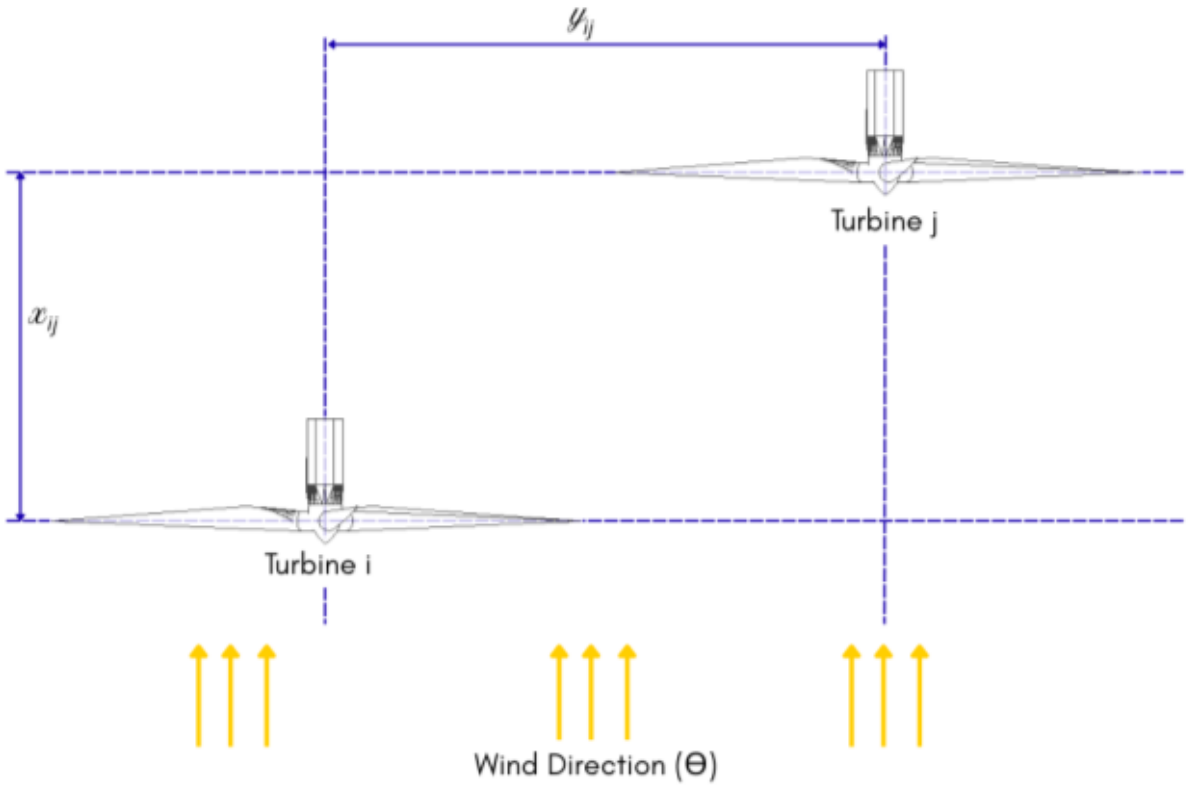

Fig. 2. Top view of the Jensen wake model.

The wind speed deficit along the wake centerline ($\Delta_c$) at downstream distance $x$ is given by (5):

$$\Delta_c(x) = \frac{1 - \sqrt{1 - C_T}}{\left(1 + 2k\frac{x}{D_i}\right)^2} \quad (5)$$

where $C_T$ is the thrust coefficient, $D_i$ is the rotor diameter of turbine $i$, and $k$ is the wake expansion coefficient. The thrust coefficient and the coefficient of power are related to the axial induction factor $\alpha$ as $C_T = 4\alpha(1 - \alpha)$ and $C_p = 4\alpha(1 - \alpha)^2$. The wake expansion coefficient $k$ depends on the turbulence intensity $I$ and is defined as $k = 0.38I + 0.004$. Higher turbulence intensities lead to faster wake recovery due to increased atmospheric mixing, while lower turbulence causes wake to persist over longer distances. Typical values of $I$ range from 0.05 – 0.1 for offshore conditions and 0.1-0.2 for onshore environments.

In the present framework, turbulence intensity is incorporated indirectly through wake expansion coefficient used in the Jensen wake model. However, detailed wake turbulence transport, turbulence induced fatigue loading, and structural degradation effects are not explicitly modeled. The objective of this work is to evaluate wake-induced velocity deficits and their influence on annual energy production and economic performance.

### *3.2. Wake Superposition for Multiple turbines*

To account for wake interactions between the turbines, the geometric overlap between an upstream turbine wake and a downstream rotor is modelled. The wake radius generated by an upstream turbine $i$ expands linearly with downstream distance $x_{ij}$ and is given by (6):

$$R_{wake}(x) = \frac{D_i}{2} + kx \quad (6)$$

To incorporate the hub height differences, the three-dimensional distance between the wake centerline and the downstream rotor's center is defined by (7) and rotor radius of the downstream turbine ($R_j$) given by (9):

$$d = \sqrt{y_{ij}^2 + \left(z_j - z_i\right)^2} \quad (7)$$

$$R_j = \frac{D_j}{2} \quad (8)$$

where $z_j - z_i$ is the hub height difference.

Wake influence is determined by calculating the overlap area between the expanding wake cross section and the downstream rotor disk. This is represented by (9):

$$\begin{aligned} &A_{overlap}\left(R_j, R_{wake}, d\right) \\ &= R_{wake}^2 Cos^{-1}\left(\frac{d^2 + R_{wake}^2 - R_j^2}{2dR_{wake}}\right) + R_i^2 Cos^{-1}\left(\frac{d^2 + R_j^2 - R_{wake}^2}{2dR_j}\right) \\ &- \frac{1}{2}\sqrt{\left(-d + R_{wake} + R_j\right)\left(d + R_{wake} - R_j\right)\left(d - R_{wake} + R_j\right)\left(d + R_{wake} + R_j\right)} \end{aligned} \quad (9)$$

Equation (9) can be simplified algebraically into three scenarios as shown below.

Scenario 1: when the distance between the centers of radius of the wake and the radius of the downstream turbine is greater than the sum of the radii, then the circles are completely separate, hence they do not overlap as seen in Fig. 3.

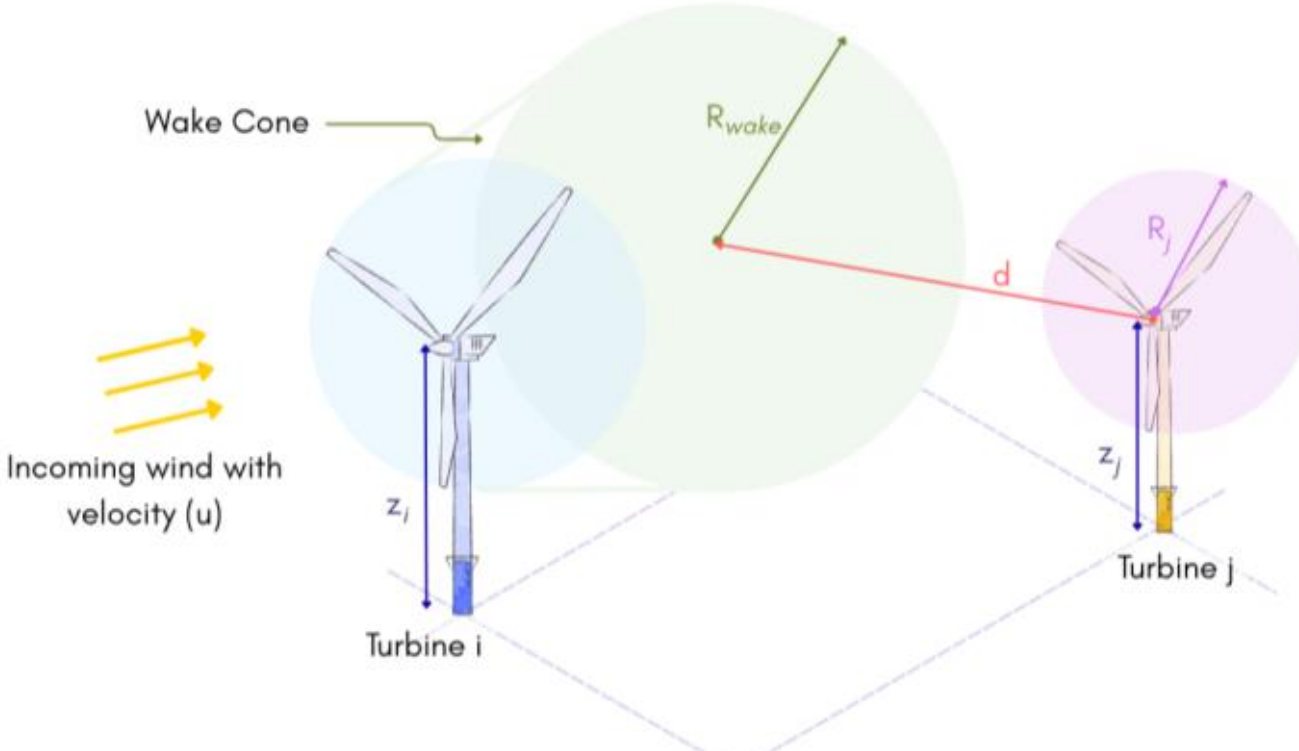

Fig. 3. Scenario 1 - No overlap between the wake cone and the downstream turbine sweep.

Scenario 2: when the distance $d$ is less than or equal to the difference of the 2 radii, it means that one circle is entirely inside the other, so the overlap area is the area of the smaller circle as shown in Fig.4.

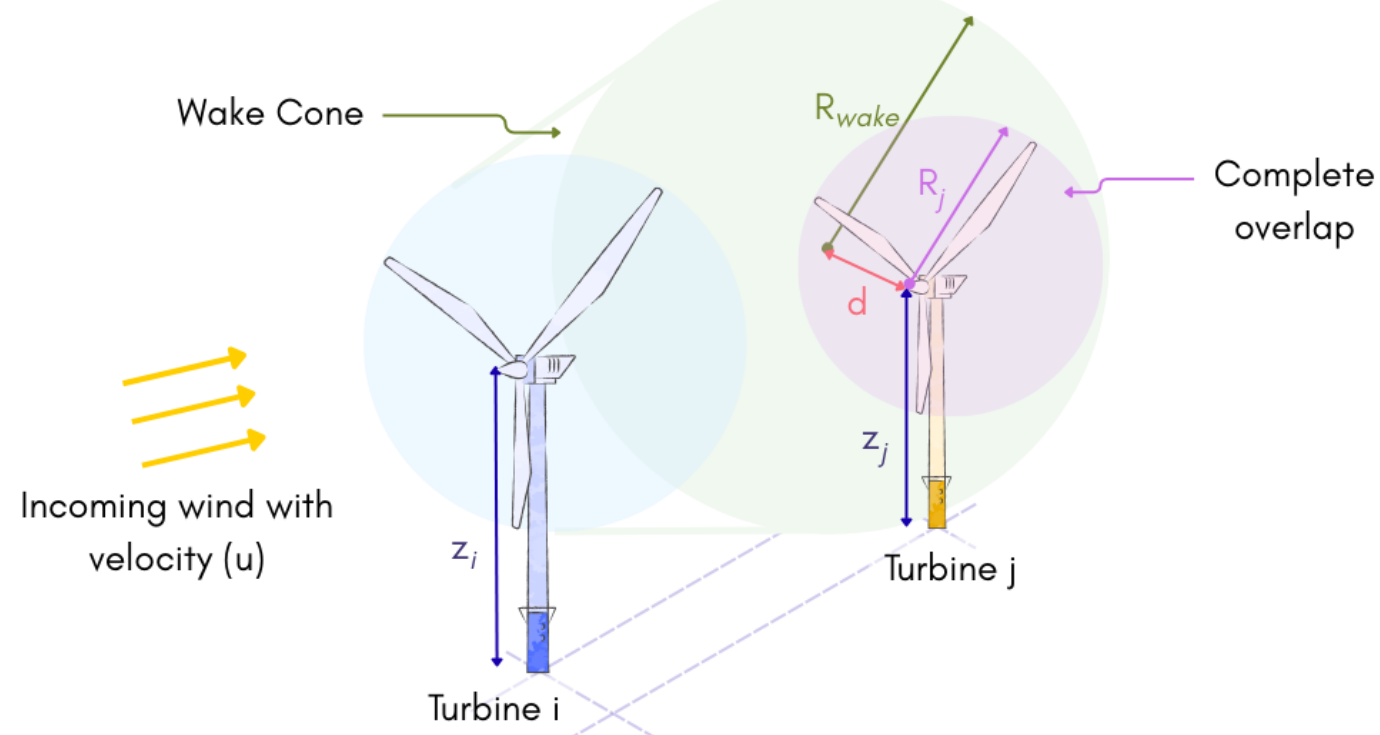

Fig. 4. Scenario 2- Full overlap of the wake cone and downstream turbine sweep.

Scenario 3: otherwise, there is a partial overlap as shown in Fig.5.

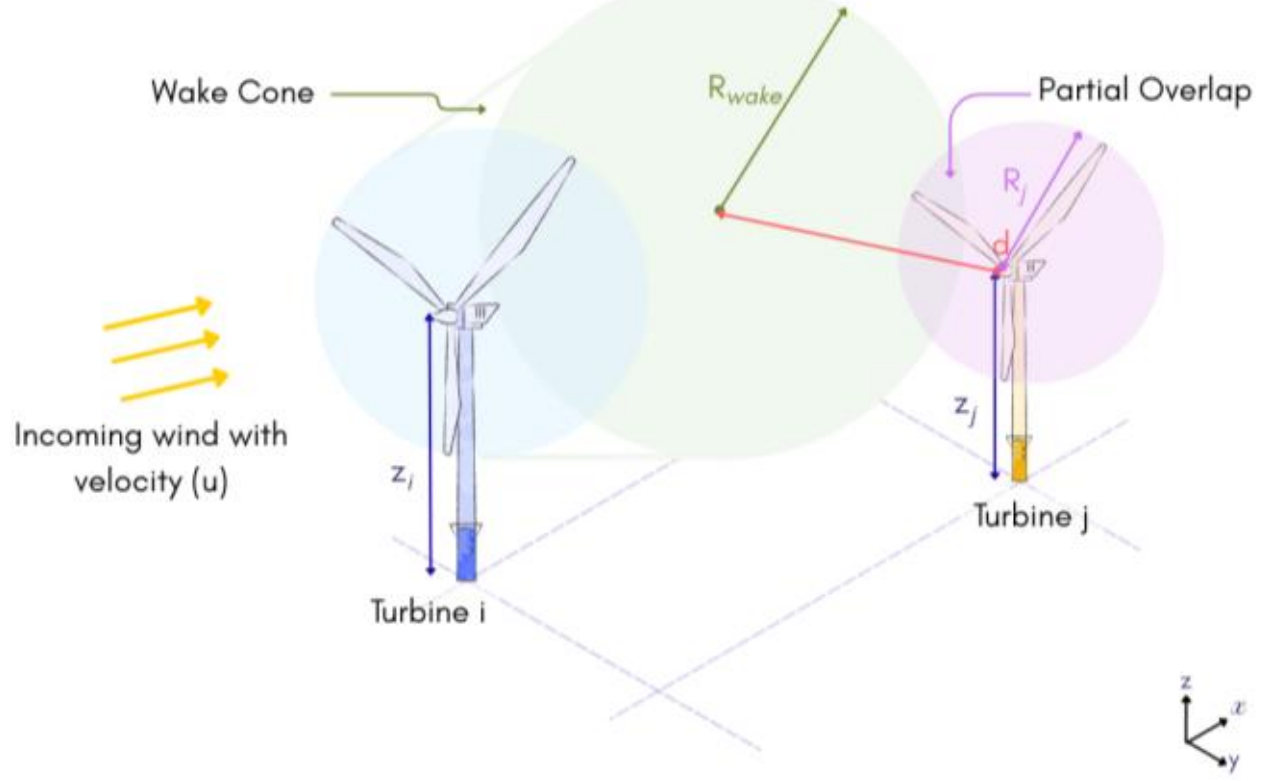

Fig. 5. Scenario 3 – Partial overlap of the wake cone and downstream turbine sweep.

Equation (9) can hence be rewritten as below:

$$A_{overlap}(R_j, R_{wake}, d) = \begin{cases} 0 & d \geq R_{wake} + R_j \\ \pi\left(\min\left(R_{Wake}, R_j\right)\right)^2 & d \leq \left|R_{wake} - R_j\right| \\ \frac{1}{2}R_{wake}^2(\Phi - \text{Sin}\Phi) + \frac{1}{2}R_j^2(\Psi - \text{Sin}\Psi) & otherwise \end{cases} \quad (10)$$

$$\Phi = 2Cos^{-1}\left(\frac{d^2 + R_{wake}^2 - R_j^2}{2dR_{wake}}\right), \qquad \Psi = Cos^{-1}\left(\frac{d^2 + R_j^2 - R_{wake}^2}{2dR_j}\right) \quad (11)$$

The resulting overlap fraction is computed using (12):

$$f_{ij} = \frac{A_{overlap}(R_j, R_{wake}, d)}{\pi R_j^2} \quad (12)$$

The velocity deficit contribution ($\delta_{ij}$) from turbine $i$ to turbine $j$ is expressed as (13):

$$\delta_{ij} = \Delta_c(x) \times f_{ij} \quad (13)$$

Multiple upstream turbines can affect a downstream turbine as shown in Fig. 6. When multiple upstream turbine wakes interact simultaneously, their combined velocity deficit is calculated using squared superposition in (14):

$$\Delta_j = \sqrt{\sum_{j \neq i} \delta_{ij}^2} \quad (14)$$

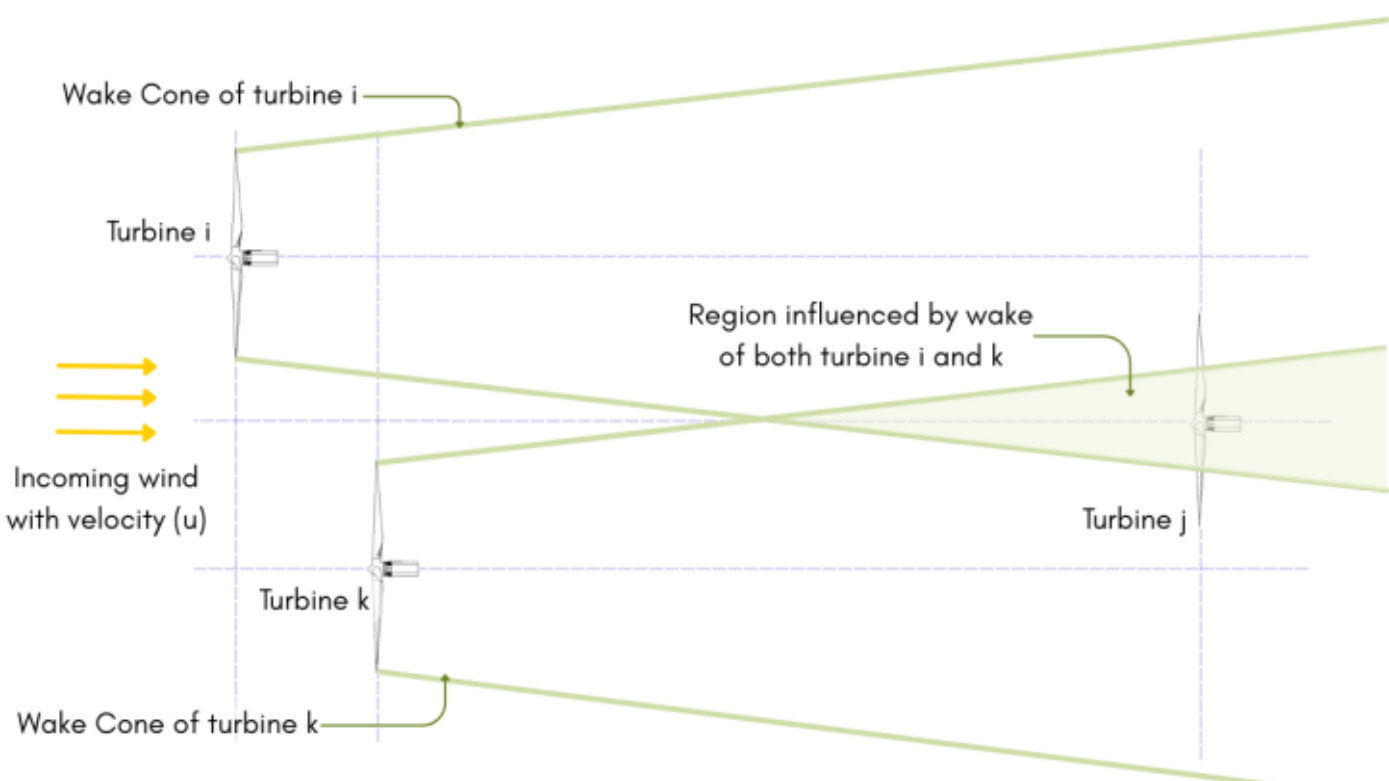

Fig. 6. Multiple upstream wakes affecting downstream turbine

The squared superposition method is commonly used in engineering wake models because linear summation may overestimate cumulative wake losses in regions with multiple overlapping wakes [48].

Finally, the effective wind speed at turbine $j$ is calculated as shown in (15):

$$v_{j,eff} = v_{j,h_2} \times \max\{0,1 - \Delta_j\} \quad (15)$$

## 4. WAKE-NET Optimization Framework

AEB is adopted as the primary optimization objective to directly reflect wind farm profitability rather than energy production alone. AEB is defined as the net economic return obtained by subtracting total investment costs from the annual production benefit (APB). In addition to energy revenue, the objective function accounts for turbine costs, land costs, and cabling costs. The resulting optimization problem is formulated through (16)-(22).

$$\max(AEB = APB - Investment) \quad (16)$$

APB is defined as the revenue generated from the wind farm's annual energy production (AEP). It is calculated as seen in (17). The AEP represents the total amount of electricity that can realistically be captured at a site and is computed by accounting for both wind speed at height $h_2$ i.e., $u = v_{h_2}$ and wind direction $\theta$ variability using long term historic data. Specifically, AEP is obtained by summing the effective power output of each turbine over all wind speeds and all direction bins, weighted by their respective occurrence probabilities ($p_\theta \times p_u$) and scaled to an annual time horizon as seen in (18):

$$APB = C_{elec} \times AEP \tag{17}$$

$$AEP = \left( N_{days} \times N_{hours} \times \sum_{\theta \in \Theta} \sum_{u \in U} \sum_{j \in T} P\left(v_{j,eff}(u,\theta)\right) \times p_\theta \times p_u \right) \tag{18}$$

The total investment cost includes the sum total of the investment costs for land, turbines, and electric cabling as calculated in (19). The land cost is proportional to the footprint area of the wind farm which is optimized by the WAKE-NET framework. The turbine cost includes both capital expenditure and operations and maintenance costs annualized over the project lifetime. The cable cost depends on the length of the cable which is optimized using the minimum spanning tree (MST) layout. Farm expansion is regulated implicitly through land and cable investment costs. The turbine cost, O&M costs, land cost, and economic parameters used in this study were adopted from published literature and industry reports [49], [50].

$$Investment = C_{land} + C_{turbine} + C_{cable} \tag{19}$$

$$C_{land} = c_{land} \times [(x_{max} - x_{min}) \times (y_{max} - y_{min})] \tag{20}$$

$$C_{turbine} = (c_{O\&M} \times T) + \left( \sum_{j=1}^{T} c_{capex,t_j} \times \frac{r}{1-(1+r)^{-N_{year}}} \right) \tag{21}$$

$$C_{cable} = \left( c_{cable} \times L_{MST} \times \frac{r}{1-(1+r)^{-N_{year}}} \right) \tag{22}$$

## 5. Case Studies

The case studies are structured into five subsections. The first subsection presents optimization results for offshore wind farm locations, followed by a second subsection focuses on onshore sites. The third subsection evaluates the impact of using multiple hub heights within a single wind farm. The fourth subsection demonstrates the importance of incorporating wake-aware model by comparing results with a benchmark wake-ignorant model. Finally, the fifth subsection analyses the influence of the number of turbines on the overall wind farm performance and profitability. The WAKE-NET framework is optimized using Genetic Algorithm with a population size of 50 and 200 generations using a fixed-generation stopping criterion. The model considers turbine of six different capacities: 8MW with hub height 90m, 11MW with hub height set at 110m, 14MW with hub height set at 125m, 16MW with hub height set at 150m, 18MW with hub height set at 160m, and 22MW with hub height set at 320m.

### *5.1. Wind Farm Sizing for Offshore Locations*

Offshore wind is stronger and more persistent than onshore wind due to reduced surface roughness and low turbulence effects. These characteristics typically enable higher energy capture and improve capacity factor which improve economic performance of offshore wind farms. Offshore turbines also tend to employ larger rotor diameters and higher hub heights, allowing greater extraction of available wind energy.

Five regions with significant offshore wind potential were selected for analysis: Alaska, Hawaii, Texas, California, and New Jersey. The wind characteristics in these regions exhibit substantial variability as illustrated in Figs. 7(a)-(e) and 8(a)-(e). The wind roses and probability distribution highlight differences in both dominant wind directions and wind speed frequencies. Wind speed bins in the wind rose are grouped into 2 m/s intervals and color-coded according to turbine operating regions: below cut-in speed (red), cut-in to rated speed (yellow), rated to cut-out speed (green), and above cut-out speed (grey).

For the Alaska site, wind speeds are predominantly westerly, with strong contributions from the west-southwest and southwest directions. The wind speed distribution indicates that wind speeds of 6-7m/s speed contributes the most frequently to energy production. Hawaii exhibits a markedly different directional pattern, with winds primarily originating from the east-northeast and east directions. The Texas offshore site demonstrates dominant winds from the south-southeast and southeast directions, while California is characterized by prevailing north-northwesterly winds. New Jersey, in contrast, shows dominant southerly wind patterns. Across all wind sites, the speed probability distribution reveals that the wind speeds occur with varying frequencies, emphasizing the importance of site-specific wind resource characterization.

These regional differences directly influence wake behavior, turbine interactions, and optimal wind farm design, thereby motivating the need for location specific optimization.

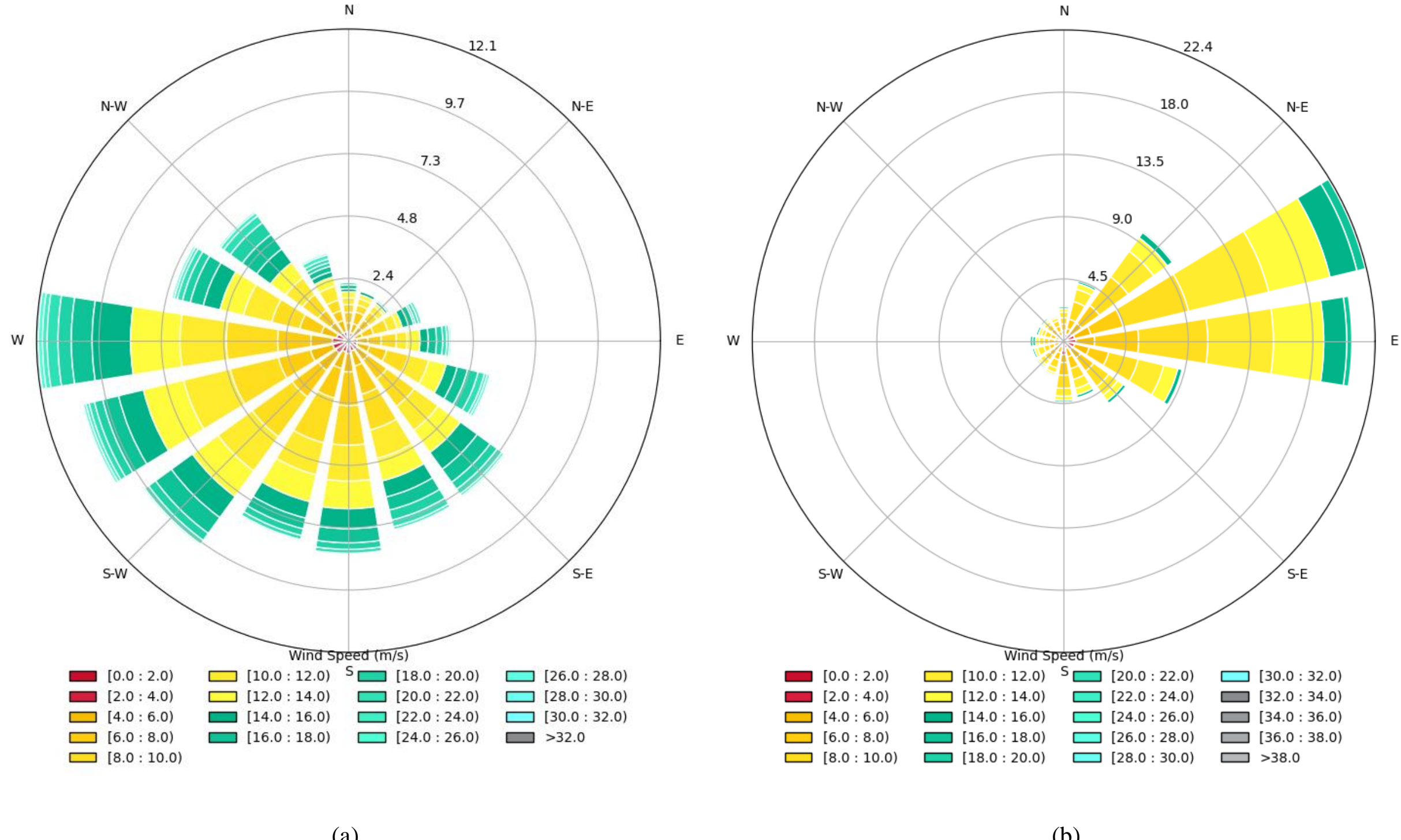

(a) (b)

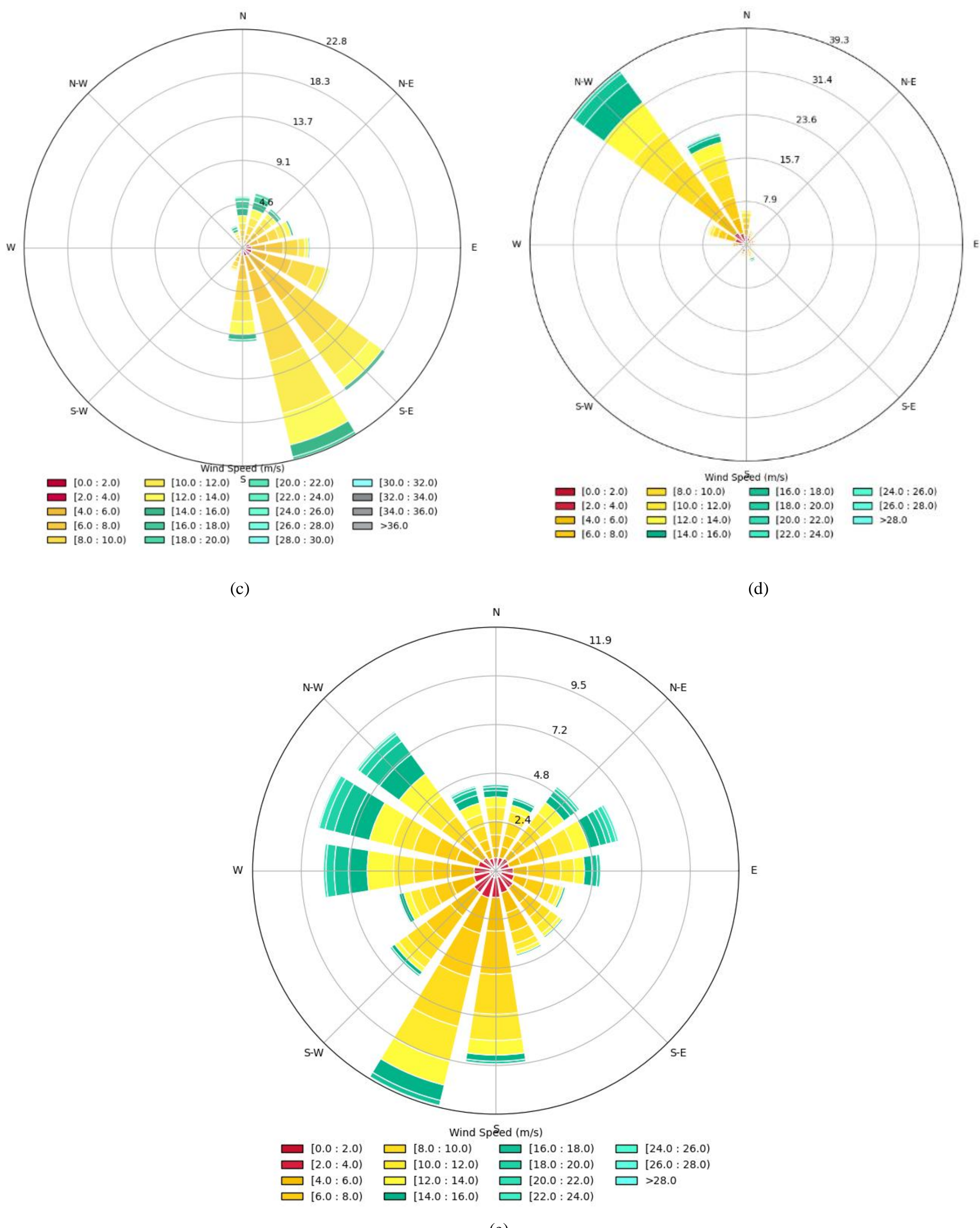

Fig. 7: Wind rose of wind blowing at 80m for offshore (a) Alaska (b) Hawaii, (c) Texas, (d) California, (e) New Jersey. Wind speed bins are grouped into 2 m/s intervals; colors indicate turbine operating regions: below cut-in speed (red), cut-in to rated speed (yellow), rated to cut-out speed (green), and above cut-out speed (grey).

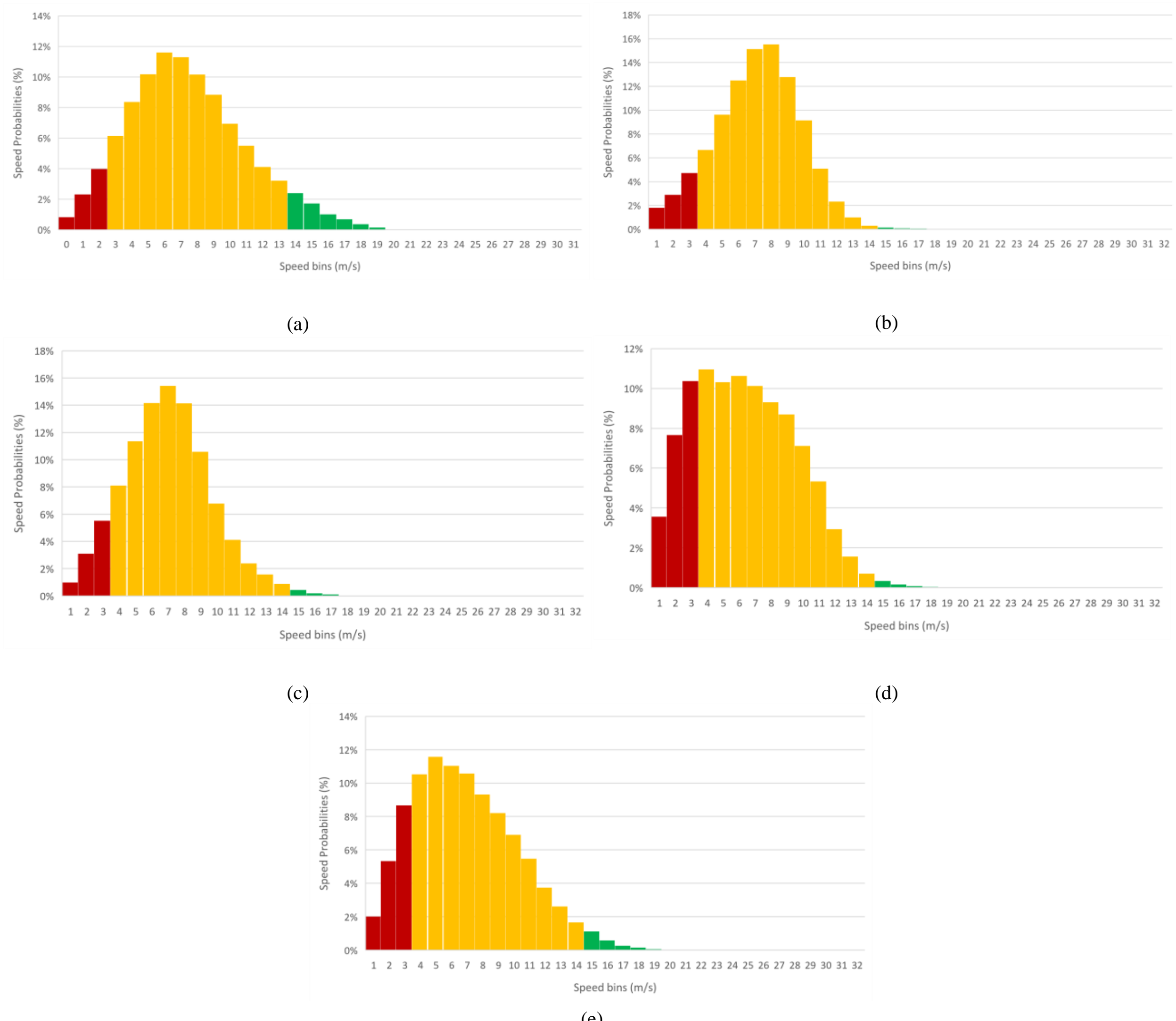

Fig. 8. Wind Speed probabilities for wind blowing at 80m for offshore (a) Alaska (b) Hawaii, (c) Texas, (d) California, (e) New Jersey. Wind speed bins are grouped into 1 m/s intervals; colors indicate turbine operating regions: below cut-in speed (red), cut-in to rated speed (yellow), rated to cut-out speed (green), and above cut-out speed (grey).

For this analysis, the number of turbines was fixed at 15 units to enable consistent comparison across various offshore regions. For visualization purposes, the wind rose representation divided the wind data into 16 directional sectors of 22.5° each, allowing a detailed depiction of prevailing wind patterns. However, to reduce computational complexity during optimization, the dataset was discretized into 12 sectors of 30° each. This simplification decreased simulation time while preserving the dominant directional characteristics of the wind resource. The WAKE-NET optimization framework assumes that all turbines are equipped with active yaw control, enabling them to continuously align with the incoming wind direction. As a result, each turbine is assumed to face the dominant wind direction within a given sector during AEP calculations.

Table 1 clearly shows that regional wind resource quality is a driving factor of offshore wind farm performance and economic viability. Although the optimized wind farms are nearly identical installed capacities, substantial variations are observed in AEP, farm capacity factor and AEB. Alaska has the highest capacity factor and AEP, resulting in an AEB, more than twice that of Hawaii and three times that of Texas.

Table 1. Optimized wind farm costs and capacities for offshore application in different regions

| ***Parameters*** | ***AK*** | ***HI*** | ***TX*** | ***CA*** | ***NJ*** |
|---|---|---|---|---|---|
| ***Avg. wind speed at 80m height (m/s)*** | 11.05 | 9.27 | 7.63 | 7.28 | 8.98 |
| ***Annual Energy Production (AEP) (GWh/year)*** | 1,549.71 | 1,197.53 | 1,139.34 | 992.78 | 1,056.32 |
| ***Annual Production Benefit (APB) (M$/year)*** | 635.38 | 490.99 | 467.13 | 407.04 | 433.09 |
| ***Farm Capacity (MW)*** | 326<br>(22MW*14)+(18MW*1) | 316<br>(22MW*12)+(18MW*2)<br>+(16MW*1) | 326<br>(22MW*14)+(18MW*1) | 320<br>(22MW*13)+(18MW*1)+<br>(16MW*1) | 306<br>(22MW*11)+(18MW*1)+<br>(16MW*2)+(14MW*1) |
| ***Capacity Factor (%)*** | 54.27 | 43.26 | 39.90 | 35.42 | 39.41 |
| ***Optimized Farm Size (sq.km)*** | 39.13 | 48.67 | 46.47 | 46.20 | 40.99 |
| ***Optimized Cable Length (km)*** | 22.56 | 22.90 | 24.53 | 24.48 | 23.61 |
| ***Annualized Land Cost (M$/year)*** | 195.63 | 243.34 | 232.36 | 230.99 | 204.96 |
| ***Annualized Cable Cost (M$/year)*** | 0.64 | 0.65 | 0.69 | 0.69 | 0.67 |
| ***Annualized Turbine Cost (M$/year) (Capex+O&M)*** | 125.86 | 122.36 | 125.86 | 123.76 | 118.87 |
| ***Annual Economic Benefit (AEB) (M$/year)*** | 313.25 | 124.64 | 108.21 | 51.59 | 108.59 |

Figures 9(a)-(e) illustrate the optimized wind farm layouts for each offshore region. For clarity, wakes corresponding to the most frequent directions are plotted: the most dominant direction is shown in red, the second in blue. Wakes are drawn at the turbines cut-out wind speed of 31.5m/s to represent maximum wake expansion.

In these offshore regions, the three most dominant wind directions are closely aligned. Because the prevailing winds originate from a relatively narrow direction band, the turbines can be arranged to minimize wake overlap. However, New Jersey experiences dominant winds from the south-southwest and west-northwest directions. As a result, the wind farm layout must be optimized to mitigate wake effects from multiple, less aligned wind sectors, which increases the design complexity. Although the average wind speed at 80m hub height is 8.98m/s, the optimized turbine mix includes several smaller rated turbines. This is because the highest wind speed probability occurs in the 4-5m/s range, meaning lower wind speeds dominate energy production over time. The elevated average wind speed is influenced by occasional high wind events exceeding 15m/s. However, these less frequent occurrences do not contribute as consistently to overall energy capture.

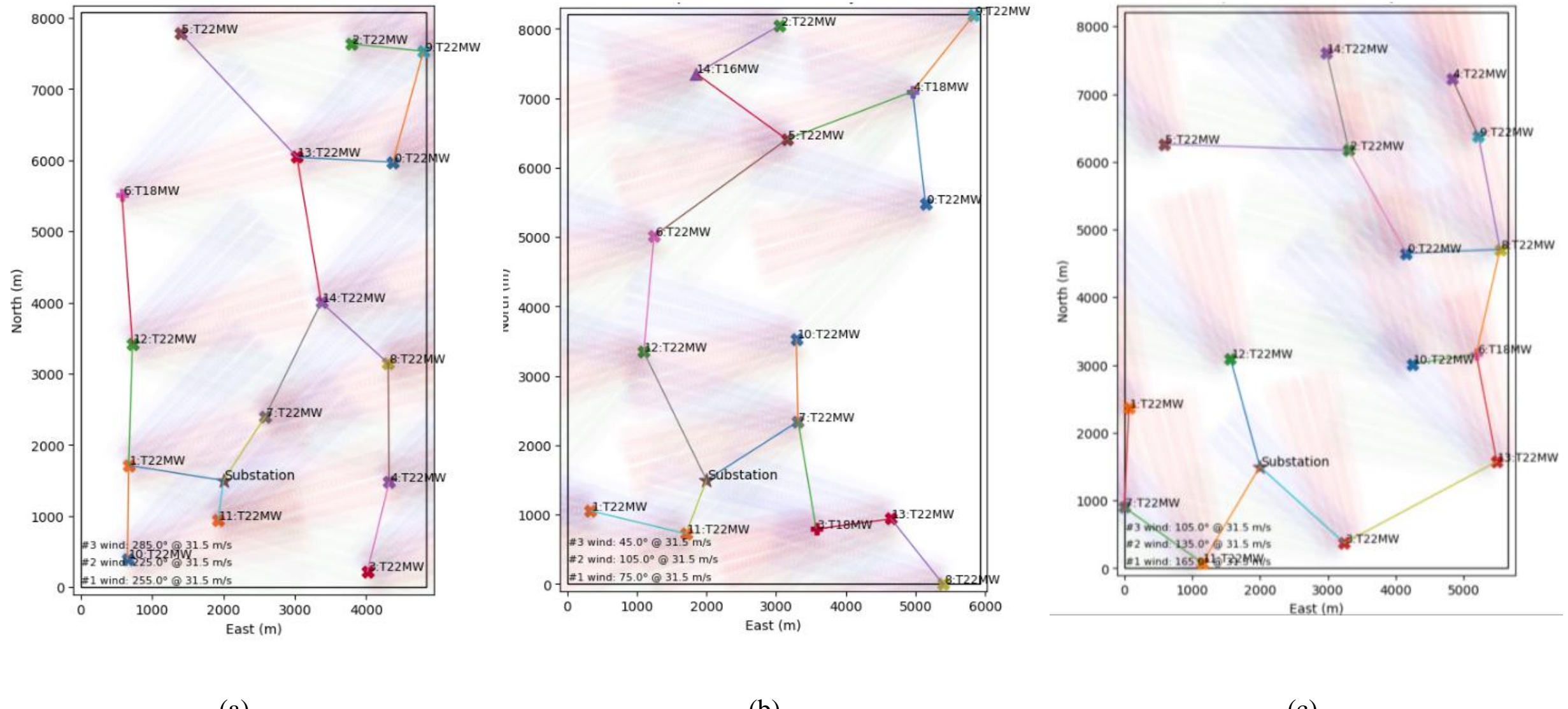

(a) (b) (c)

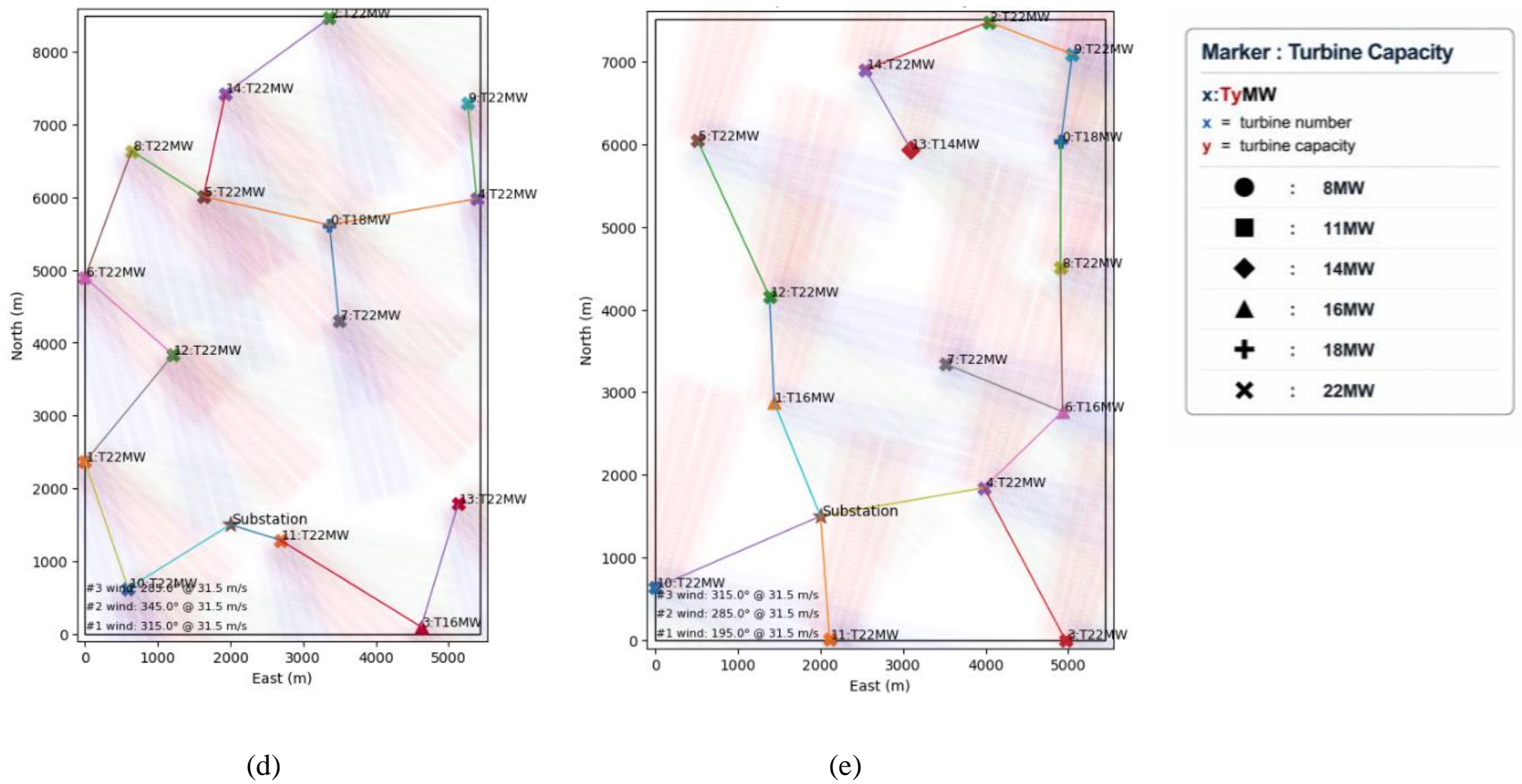

(d) (e)

Fig. 9. Wind farm optimized for offshore (a) Alaska (b) Hawaii, (c) Texas, (d) California, (e) New Jersey, with 15 turbines each farm. Turbine ratings and corresponding hub heights are 8 MW (90 m), 11 MW (110 m), 14 MW (125 m), 16 MW (150 m), 18 MW (160 m), and 22 MW (320 m).

## *5.2. Wind Farm Sizing for Onshore Locations*

Four onshore locations were selected for analysis: Texas, Kansas, Wyoming, and Indiana. For the onshore optimization scenarios, simplified flat-terrain conditions were assumed and localized microscale topographic effects were not explicitly modeled. In general, onshore wind resources are weaker and more variable than offshore winds due to increased surface roughness and terrain effects. As a result, careful site selection becomes especially important for ensuring economic viability.

The Great Plains region of the U.S., which includes Texas and Kansas, is well known for its strong and persistent winds. These favorable conditions are primarily driven by significant temperature gradients and relatively flat terrain, which allows air masses to move with minimal obstruction. Wyoming also exhibits strong wind resources due to its unique topography that creates a natural air tunnel effect that accelerates winds through mountain passes and valleys. In contrast, the eastern U.S. typically experiences weaker and more variable onshore winds, making locations such as Indiana comparatively less favorable for large scale wind development.

Figures 10 (a)-(d) and 11(a)-(d) present the wind roses and wind speed probability distributions for the selected onshore sites. Texas and Kansas show dominant wind speeds primarily distribution for the selected onshore sites. Texas and Kansas show dominant winds primarily for the southern sectors, consistent with Great Plains wind patterns. Wyoming, however, is characterized by strong west-southwesterly winds. While lower wind speeds have slightly higher occurrence probability across many sites, Wyoming and the Great Plains locations also experience frequent high speed wind events. Indiana, in contrast, is dominated by lower wind speeds, which limits its overall energy production potential.

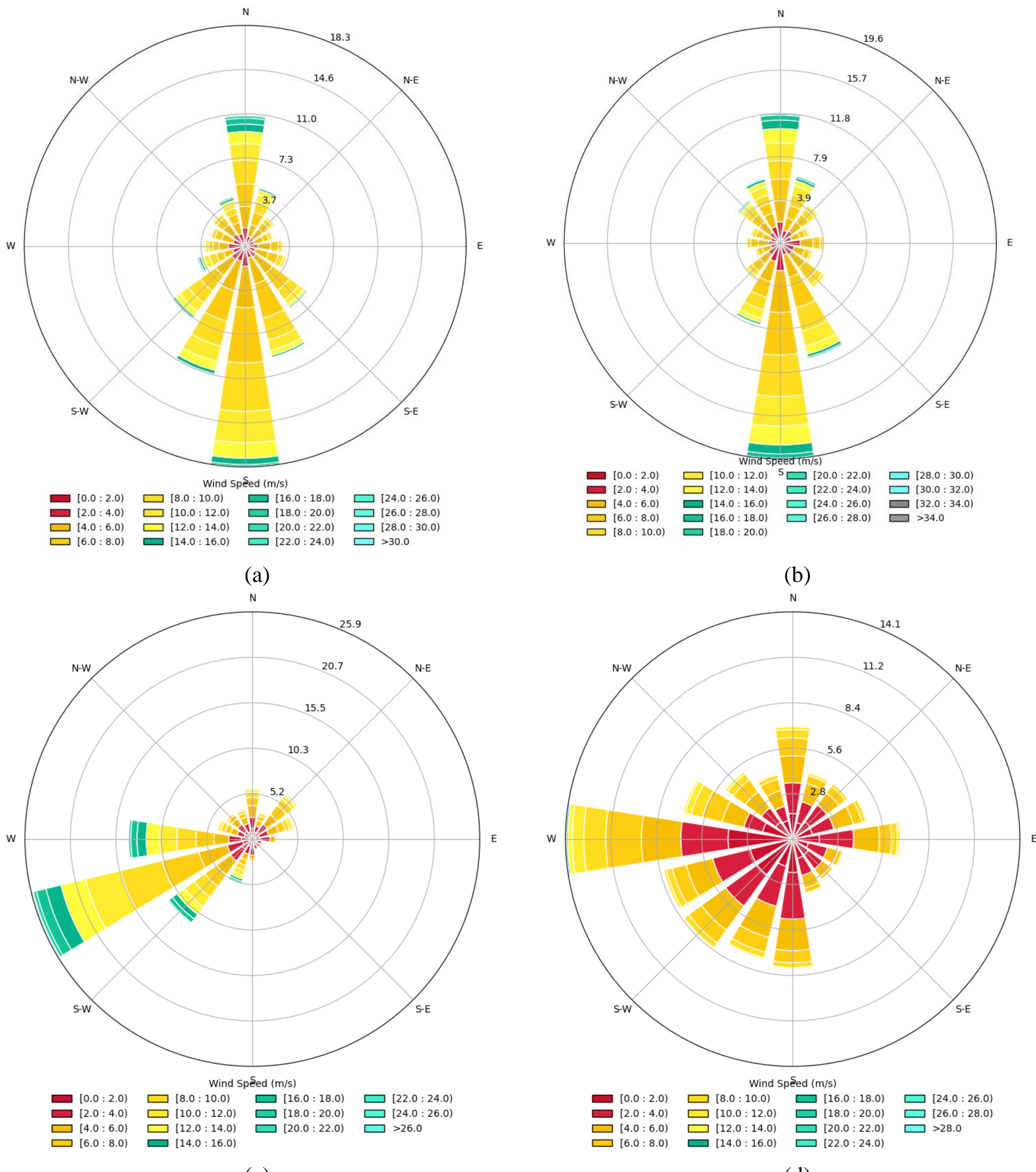

Fig. 10: Wind rose of wind blowing at 80m for onshore (a) Texas, (b) Kansas, (c) Wyoming, (d) Indiana. Wind speed bins are grouped into 2 m/s intervals; colors indicate turbine operating regions: below cut-in speed (red), cut-in to rated speed (yellow), rated to cut-out speed (green), and above cut-out speed (grey).

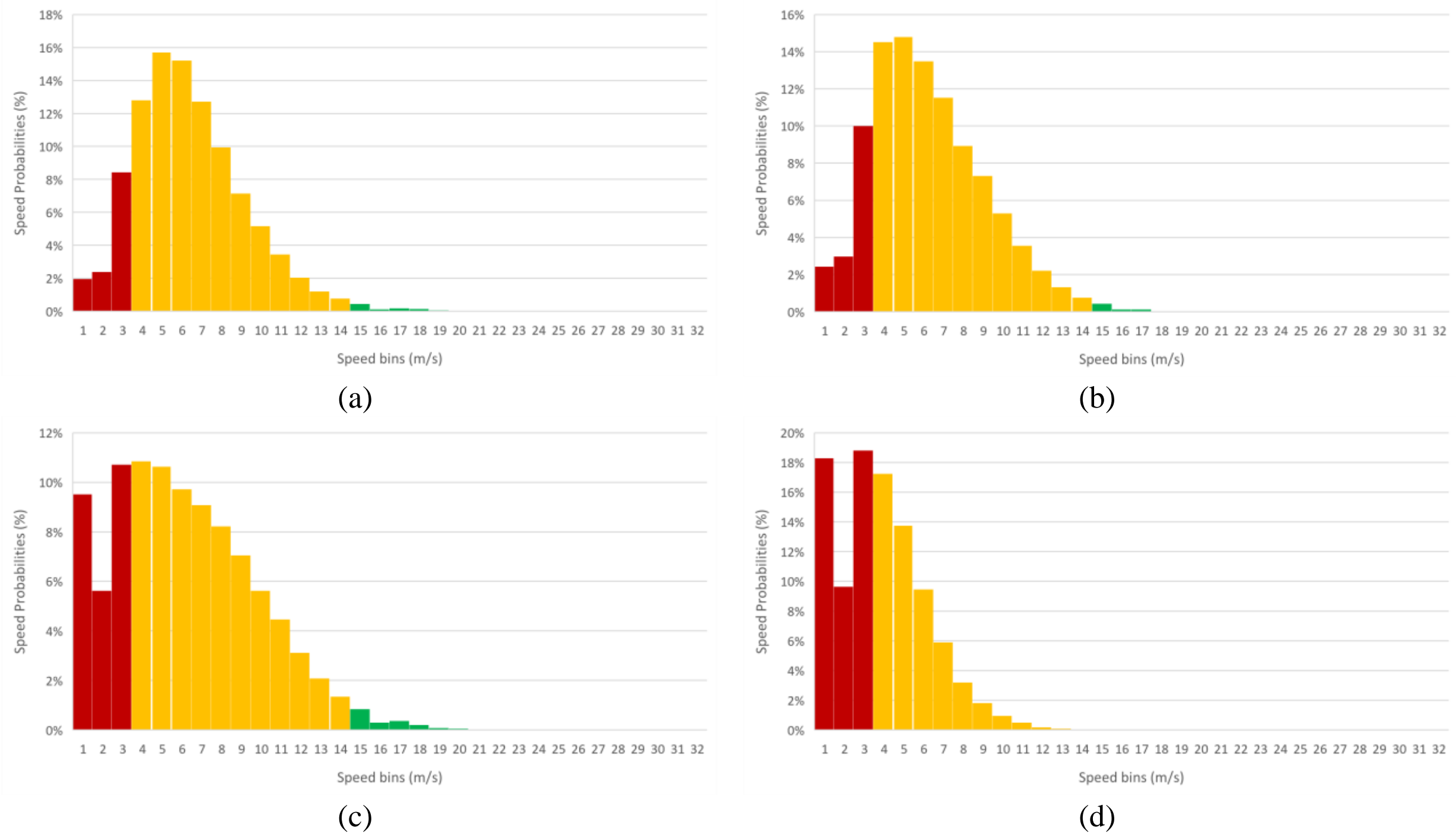

Fig. 11. Wind Speed probabilities for wind blowing at 80m for onshore (a) Texas, (b) Kansas, (c) Wyoming, (d) Indiana. Wind speed bins are grouped into 1 m/s intervals; colors indicate turbine operating regions: below cut-in speed (red), cut-in to rated speed (yellow), rated to cut-out speed (green), and above cut-out speed (grey).

The total number of turbines is fixed to 15 units for each optimization scenario. Table 2 and Fig. 12(a)-(d) present the optimized wind farm capacities, costs, and layout configurations for the selected onshore locations. The results reflect the strong influence of regional wind resource quality on the farm performance.

Table 2. Optimized wind farm costs and capacities for onshore application in different regions

| *Parameters* | *TX* | *KS* | *WY* | *IN* |
|---|---|---|---|---|
| ***Avg. wind speed at 80m height (m/s)*** | 7.28 | 7.09 | 6.96 | 4.02 |
| ***Annual Energy Production (AEP) (GWh/year)*** | 823.177 | 803.78 | 893.602 | 193.267 |
| ***Annual Production Benefit (APB) (M$/year)*** | 337.50 | 329.55 | 366.38 | 79.24 |
| ***Farm Capacity (MW)*** | 293<br>(22MW*11)+(18MW*1)+(14MW*1)+(11MW*1)+(8MW*1) | 298<br>(22MW*9)+(18MW*2)+(16MW*4) | 302<br>(22MW*12)+(16MW*1)+(11MW*2) | 227<br>(22MW*5)+(18MW*1)+(16MW*3)+(11MW*1)+(8MW*5) |
| ***Capacity Factor (%)*** | 32.07 | 30.79 | 33.78 | 9.72 |
| ***Optimized Farm Size (sq.km)*** | 40.45 | 36.39 | 44.25 | 37.11 |
| ***Optimized Cable Length (km)*** | 23.53 | 19.23 | 22.47 | 21.01 |
| ***Annualized Land Cost (M$/year)*** | 202.26 | 181.93 | 221.23 | 185.57 |
| ***Annualized Cable Cost (M$/year)*** | 0.67 | 0.54 | 0.64 | 0.60 |
| ***Annualized Turbine Cost (M$/year) (Capex+O&M)*** | 114.33 | 116.08 | 117.47 | 91.28 |
| ***Annual Economic Benefit (AEB) (M$/year)*** | 20.25 | 30.99 | 27.04 | -198.21 |

Texas has a higher average wind speed and has a higher proportion of the wind speed above the 3m/s, resulting in a higher AEP despite selecting a slightly smaller installed capacity. Wyoming achieves the highest AEP amongst the onshore locations and supports a larger optimized farm capacity as it has some of the highest onshore wind speeds unique to its geographic location. All three of these regions achieve farm capacity factor above 30%, which is consistent with wind farms in these locations.

In contrast, the Indiana site performs significantly worse due to its lower wind speeds. The optimizer selects a smaller installed capacity. However, even with this adjustment, the resulting farm capacity factor remains below 10%. Consequently, the AEB is negative, indicating that under the assumed cost conditions, the project would not be economically viable without financial support. This result does not imply that wind development in Indiana is impossible, but rather that such projects would likely require policy incentives or subsidies to achieve economic feasibility.

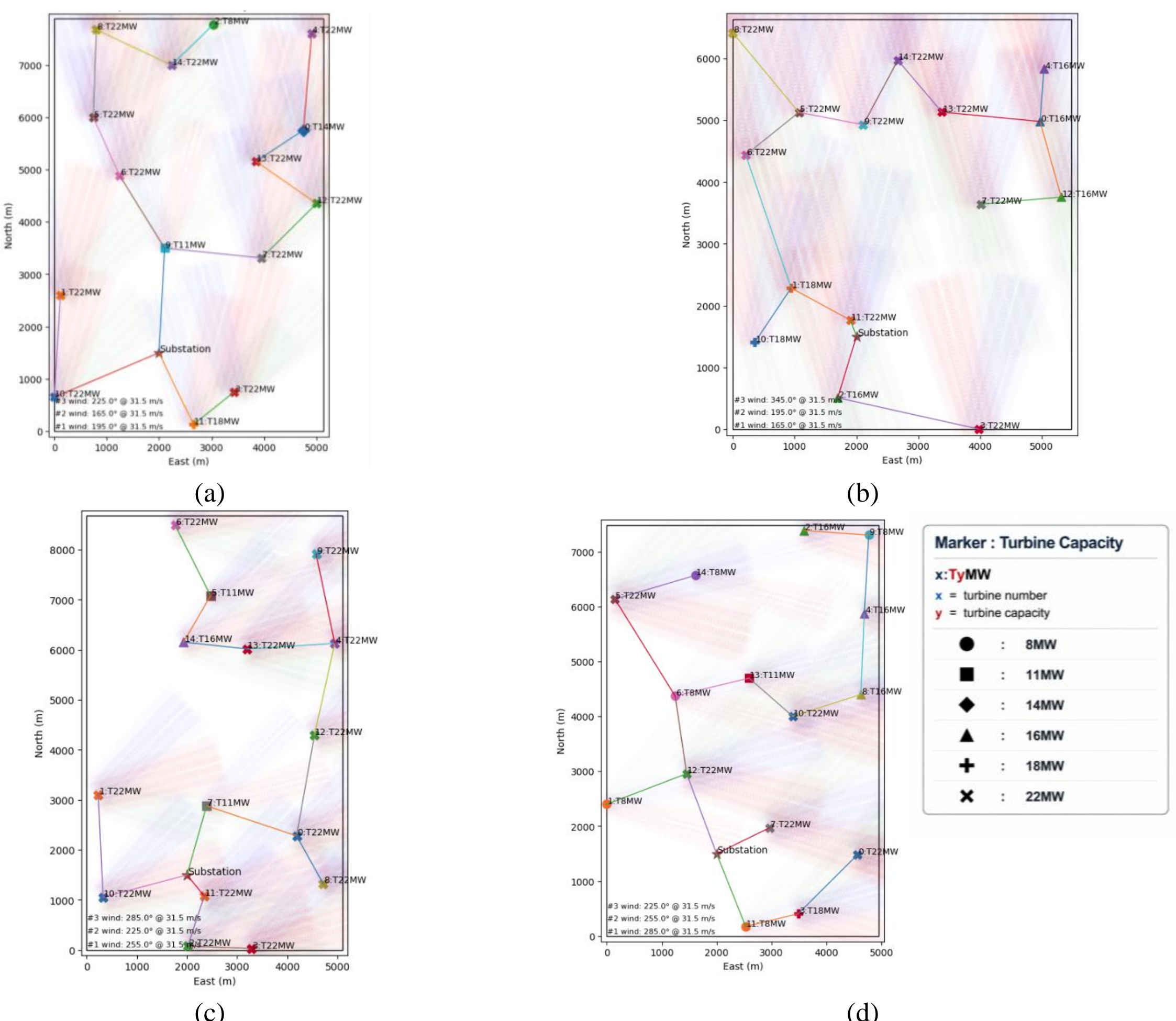

Fig. 12. Wind farm optimized for onshore (a) Texas, (b) Kansas, (c) Wyoming, (d) Indiana, with 15 turbines each farm. Turbine ratings and corresponding hub heights are 8 MW (90 m), 11 MW (110 m), 14 MW (125 m), 16 MW (150 m), 18 MW (160 m), and 22 MW (320 m).

### *5.3. Multiple Hub Heights in a Single Wind Farm*

In this section, we investigate whether deploying turbines at multiple hub heights can reduce wake interactions and improve overall wind farm performance. To evaluate this hypothesis, three hub height configurations were compared for offshore Alaska and offshore Hawaii. To isolate the economic impact of wake mitigation and hub-height diversification, turbine cost parameters were assumed constant for each turbine rating across all scenarios. Consequently, additional structural and foundation costs associated with taller towers were not explicitly modeled in this comparative analysis.

The three cases are defined as follows:

Case 1 (six hub heights): 8MW with hub height 90m, 11MW with hub height set at 110m, 14MW with hub height set at 125m, 16MW with hub height set at 150m, 18MW with hub height set at 160m, and 22MW with hub height set at 320m.

Case 2 (two hub height levels): 8MW, 11MW and 14MW turbines have hub height 125m. 16MW, 18MW and 22MW turbines have hub height 320m.

Case 3 (single hub height): All capacities of turbines are installed at 320m.

Tables 3 and 4, along with Fig. 13(a)-(b) and 14(a)-(b), present the results for a 15-turbine wind farm in Alaska and Hawaii, respectively.

Although higher hub heights generally experience stronger wind speeds, the results indicate that the additional energy gained is partially offset by increased wake interactions. Taller turbines require larger land spacing to mitigate wake overlap, leading to increased farm areas. In some cases, the optimizer reduces the installed capacity to control wake losses. Consequently, increases in hub height do not necessarily translate to higher AEP or AEB.

Similar trends are observed in both Alaska and Hawaii suggesting that while multiple hub heights can improve vertical wind resource utilization, their economic advantage depends on the balance between wind shear benefits and wake-induced losses.

Table 3. Comparison of multiple hub heights for wind farms in Alaska

| ***Parameters*** | ***Case 1*** | ***Case 2*** | ***Case 3*** |
|---|---|---|---|
| ***Annual Energy Production (AEP) (GWh/year)*** | 1,549.71 | 1,558.21 | 1,465.55 |
| ***Annual Production Benefit (APB) (M$/year)*** | 635.38 | 638.87 | 600.88 |
| ***Farm Capacity (MW)*** | 326<br>(22MW*14)+(18MW*1) | 324<br>(22MW*14)+(16MW*1) | 309<br>(22MW*12)+(18MW*1)+(16MW*1)+(11MW*1) |
| ***Capacity Factor (%)*** | 54.27 | 54.90 | 54.14 |
| ***Optimized Farm Size (sq.km)*** | 39.13 | 45.28 | 44.95 |
| ***Optimized Cable Length (km)*** | 22.56 | 22.36 | 22.84 |
| ***Annualized Land Cost (M$/year)*** | 195.63 | 226.39 | 224.77 |
| ***Annualized Cable Cost (M$/year)*** | 0.64 | 0.63 | 0.65 |
| ***Annualized Turbine Cost (M$/year) (Capex+O&M)*** | 125.86 | 125.16 | 119.92 |
| ***Annual Economic Benefit (AEB) (M$/year)*** | 313.25 | 286.69 | 255.54 |

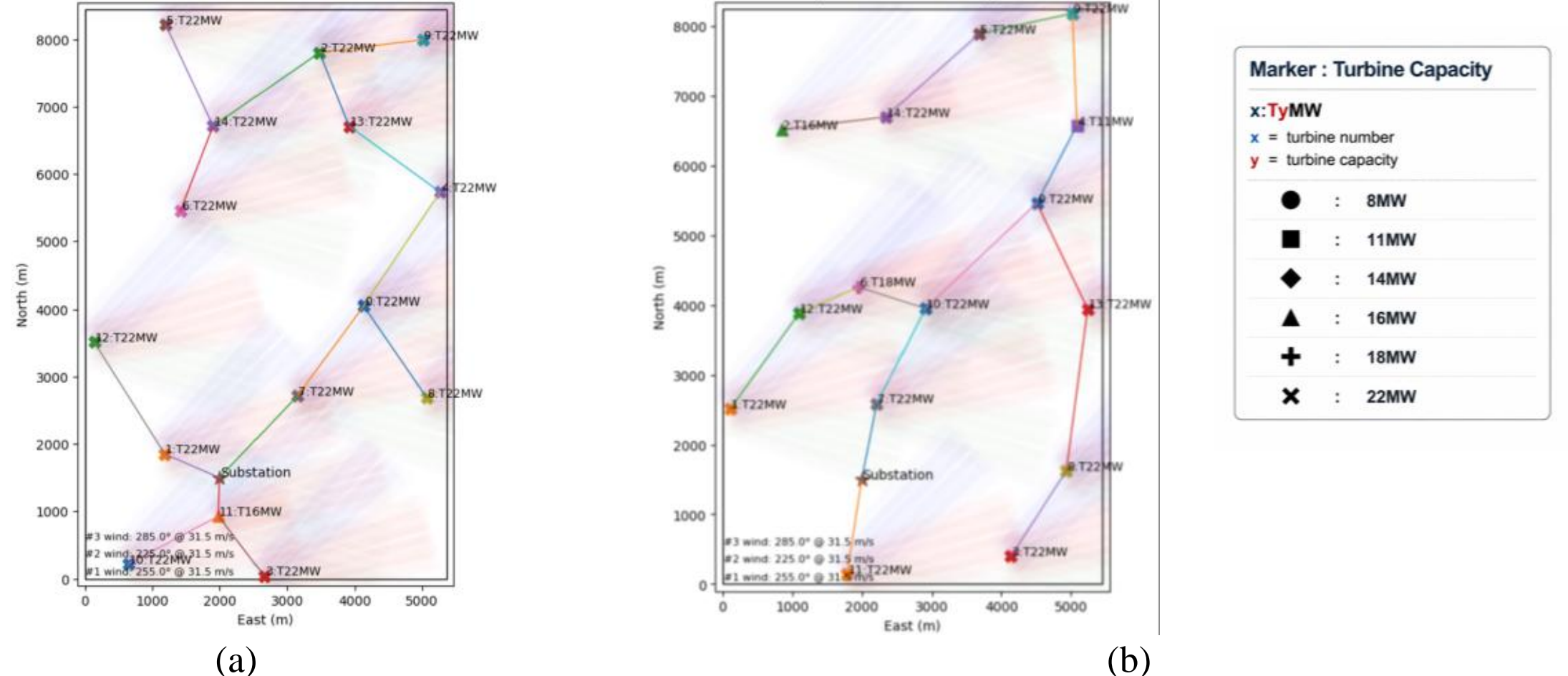

(a) (b)

Fig. 13. Wind farm optimized for offshore Alaska when hub heights are set as per (a) Case 2 (dual hub-height configuration), where 8–14 MW turbines are installed at 125 m and 16–22 MW turbines at 320 m; and (b) Case 3 (single hub-height configuration), where all turbines are installed at 320 m, for 15 turbines in each farm.

Table 4. Comparison of multiple hub heights for wind farms in Hawaii

| ***Parameters*** | ***Case 1*** | ***Case 2*** | ***Case 3*** |
|---|---|---|---|
| ***Annual Energy Production (AEP) (GWh/year)*** | 1,197.53 | 1,161.00 | 1,106.82 |
| ***Annual Production Benefit (APB) (M$/year)*** | 490.99 | 476.01 | 453.80 |
| ***Farm Capacity (MW)*** | 316<br>(22MW*12)+(18MW*2)+(16MW*1) | 308<br>(22MW*11)+(18MW*2)+(16MW*1)+(14MW*1) | 298<br>(22MW*9)+(18MW*3)+(16MW*2)+(14MW*1) |
| ***Capacity Factor (%)*** | 43.26 | 43.03 | 42.40 |
| ***Optimized Farm Size (sq.km)*** | 48.67 | 50.33 | 45.85 |
| ***Optimized Cable Length (km)*** | 22.90 | 22.78 | 20.89 |
| ***Annualized Land Cost (M$/year)*** | 243.34 | 251.64 | 229.25 |
| ***Annualized Cable Cost (M$/year)*** | 0.65 | 0.65 | 0.59 |
| ***Annualized Turbine Cost (M$/year) (Capex+O&M)*** | 122.36 | 119.57 | 116.08 |
| ***Annual Economic Benefit (AEB) (M$/year)*** | 124.64 | 104.15 | 107.88 |

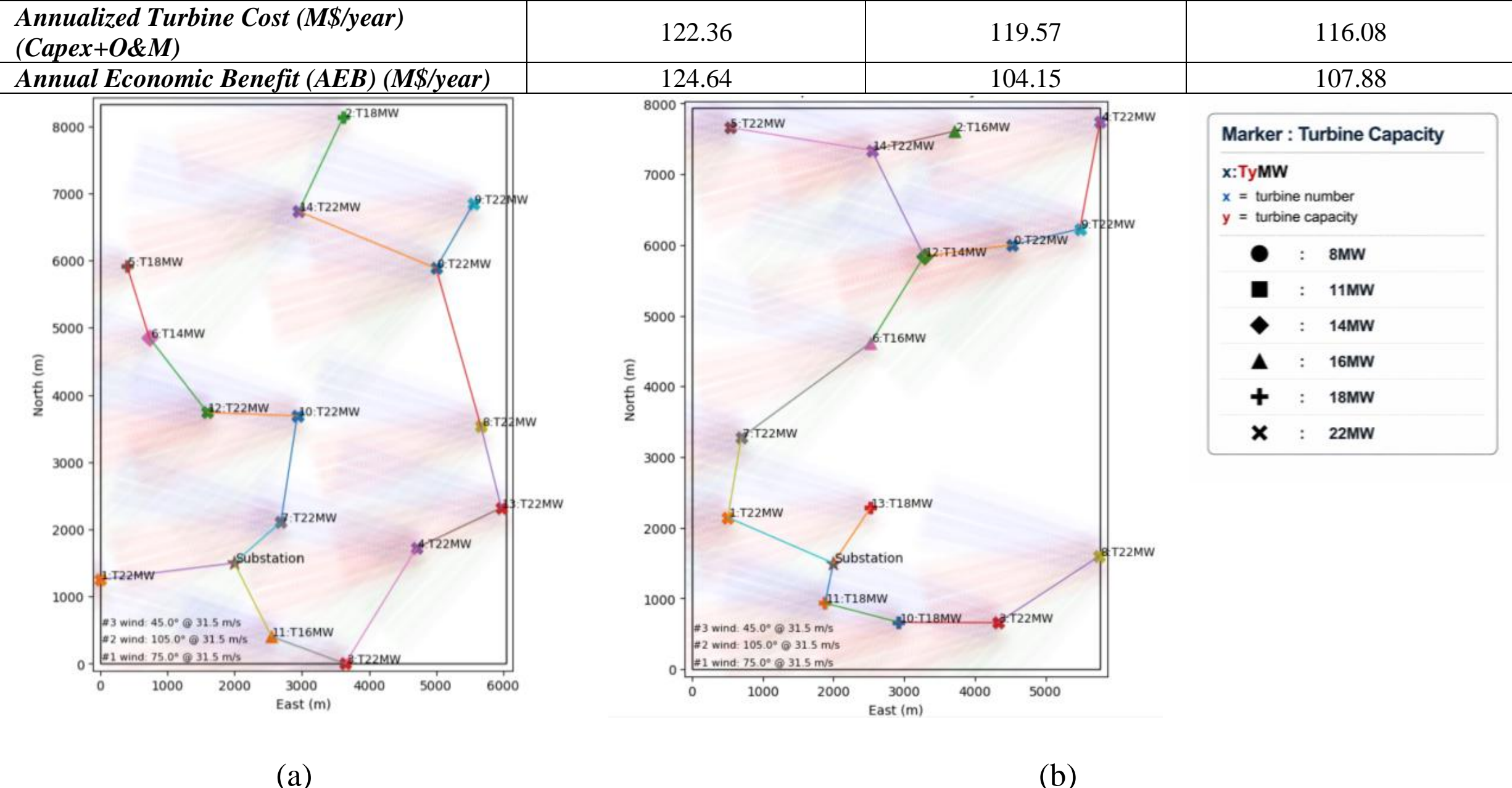

Fig. 14. Wind farm optimized for offshore Hawaii when hub heights are set as per (a) Case 2 (dual hub-height configuration), where 8–14 MW turbines are installed at 125 m and 16–22 MW turbines at 320 m; and (b) Case 3 (single hub-height configuration), where all turbines are installed at 320 m for 15 turbines in each farm.

## *5.4. Comparison of Wake-Aware and Wake-Ignorant Models*

This comparative analysis was conducted for the offshore California site using 15 turbines within a constrained area of 10km x 10km. Two optimization approaches were evaluated. The wake-aware model incorporated wake interactions during layout optimization, while the benchmark wake-ignorant model neglected wake effects entirely. The wake-ignorant model was evaluated in two stages: first, the layout was optimized under identical spatial and economic constraints, but without considering wake induced velocity deficits; second, a post-analysis was performed in which wake effects were applied to the resulting layout to assess its realistic performance. The results are summarized in table 5.

Table 5. Comparison of wake-aware and wake-ignorant model for California

| ***Parameters*** | ***Wake-Aware*** | ***Wake-Ignorant*** | |
|---|---|---|---|
| | | No wake | Post-optimization wake consideration |
| ***Annual Energy Production (AEP) (GWh/year)*** | 992.78 | 972.98 | 926.50 |
| ***Annual Production Benefit (APB) (M$/year)*** | 407.04 | 398.92 | 379.87 |
| ***Farm Capacity (MW)*** | 320<br>(22MW*13)+(18MW*1)+ (16MW*1) | 304<br>(22MW*10)+(18MW*2)+(16MW*3) | |
| ***Capacity Factor (%)*** | 35.42 | 36.54 | 34.79 |
| ***Optimized Farm Size (sq.km)*** | 46.20 | 44.87 | |
| ***Optimized Cable Length (km)*** | 24.48 | 19.99 | |
| ***Annualized Land Cost (M$/year)*** | 230.99 | 224.33 | |
| ***Annualized Cable Cost (M$/year)*** | 0.69 | 0.57 | |
| ***Annualized Turbine Cost (M$/year) (Capex+O&M)*** | 123.76 | 118.17 | |
| ***Annual Economic Benefit (AEB) (M$/year)*** | 51.59 | 55.85 | 36.79 |

Although both models were applied to the same site and constraints, the developed distinctly different wind farm configurations. The wake-aware model designed a 320MW farm, while the wake-ignorant model produced a slightly smaller 304MW layout. When wake effects were ignored, the wake-ignorant model predicted an AEP of 972.98GWh annually. However, after applying wake losses in the post analysis, its AEP decreased by 46.48GWh/year due to significant wake interference, as illustrated in Fig. 15. A similar observation is made in the APB. The wake-aware layout optimized by the WAKE-NET framework achieves an AEB that is 50.59% greater per annum than the wake-ignorant layout.

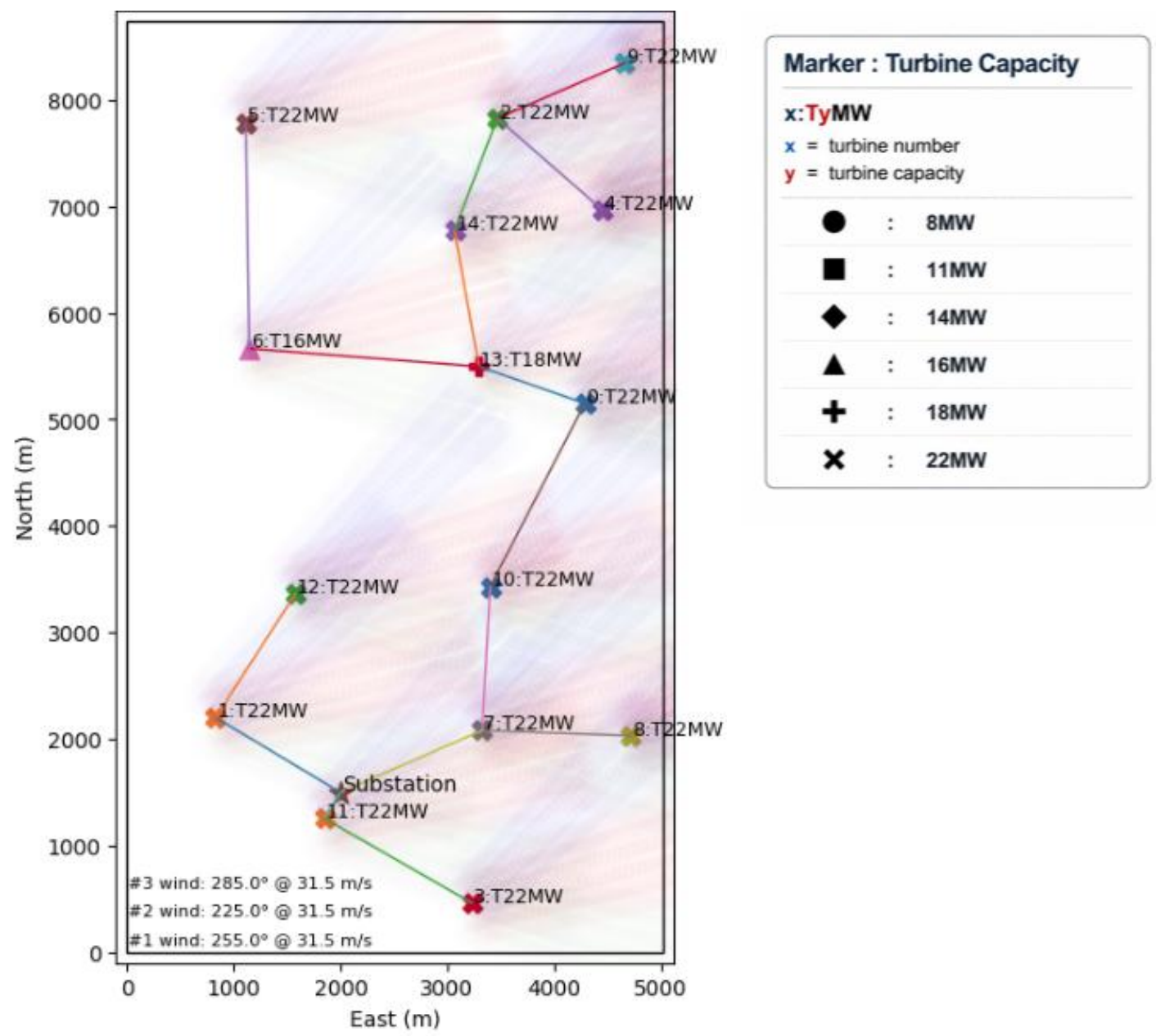

Fig. 15. Wind farm layout for offshore California proposed by the wake-ignorant model. Turbine ratings and corresponding hub heights are 8 MW (90 m), 11 MW (110 m), 14 MW (125 m), 16 MW (150 m), 18 MW (160 m), and 22 MW (320 m).

### *5.5. Performance Sensitivity to Wind Farm Scaling*

If land area were unconstrained, increasing the number of turbines would theoretically yield unbounded revenue due to the continuous growth in installed capacity. In practice, however, wind farm deployment is inherently limited by spatial constraints. The previous case study demonstrated that incorporating turbines with multiple hub heights can improve special utilization and reduce the required farm area. Nevertheless, even with this approach, turbine density cannot be increased arbitrarily without incurring performance penalties.

To examine this tradeoff, a sensitivity analysis was performed by constraining the available wind farm area to 145 sq.km, comparable to the scale of the largest wind farms currently operating in the U.S. For the offshore Alaska site, the sensitivity analysis evaluates the impact of turbine count on farm performance metrics as seen in Fig. 16.

As the number of turbines increases the overall farm capacity factor exhibits a gradual decline. This trend is expected as higher turbine densities increase wake interactions resulting in larger velocity deficits and deduced energy capacity by downstream turbines. A similar pattern is observed for the AEB per turbine, which decreases with increased turbine count. Although additional turbines increase the total installed capacity, the marginal energy contribution of each turbine diminishes due to wake-induced losses. Consequently, while the total AEB initially rises with turbine count, it reaches a maximum and subsequently declines. For this case, the peak occurs at around twenty-six turbines. Beyond this point, the combined effects of intensified wake losses and increased investment costs outweigh the gain in APB. The optimized wind farm area increases with turbine count at lower deployment densities by eventually approaches a plateau.

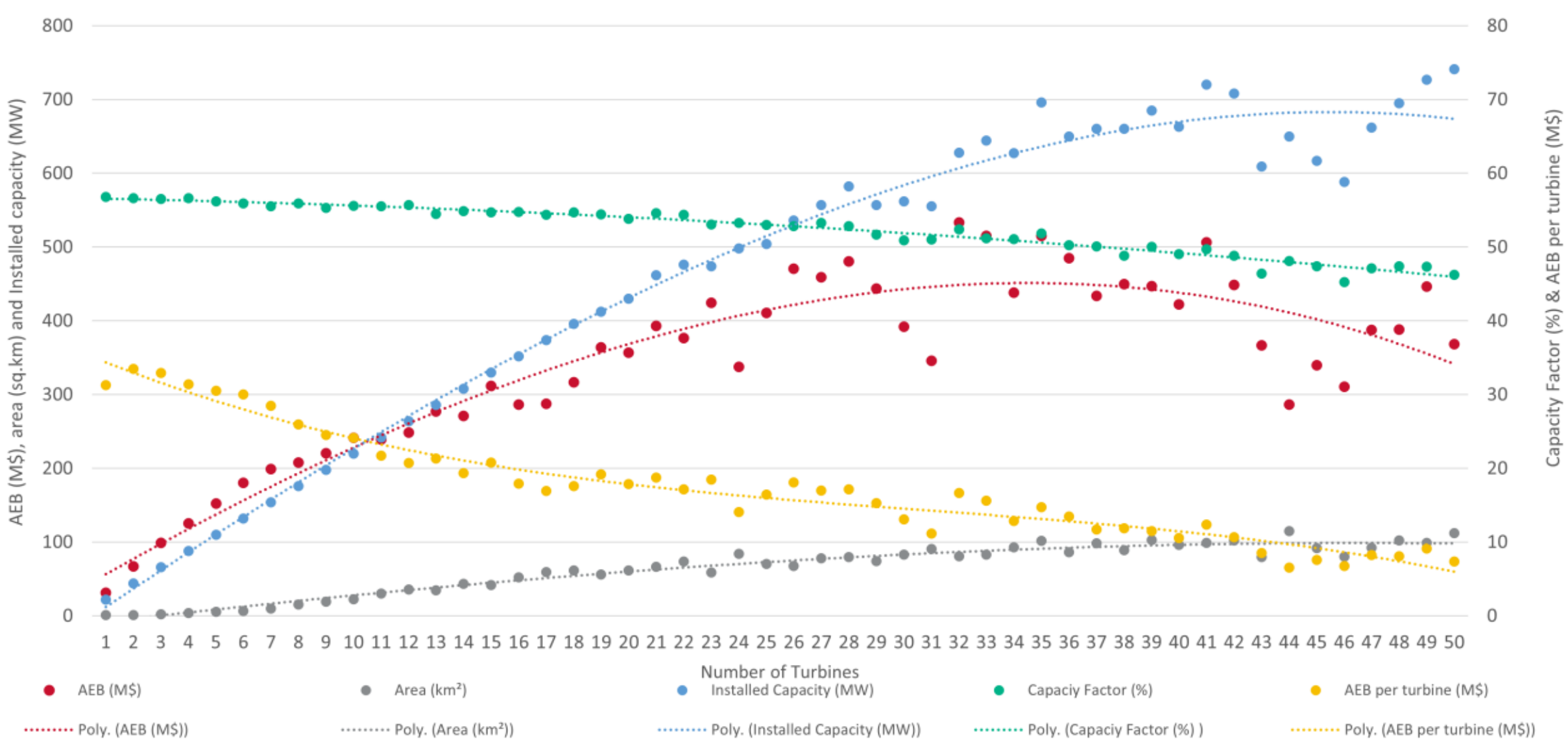

Fig. 16. Impact of wind farm scaling on farm performance

These results emphasize that wind farm optimization should target an economically optimal turbine count rather than maximum turbine deployment as excessive turbine density degrades both capacity factor and profitability. Another observation is that computational cost increased significantly with turbine count due to the growth in wake interaction evaluations. Representative runtimes ranged from approximately 30 minutes for 15-turbine scenarios to approximately 30 hours for 50-turbine scenarios on a standard workstation.

## 6. Conclusions

This study developed an integrated wake-aware optimization framework for wind farm layout design that maximizes AEB while incorporating aerodynamic wake interactions, turbine characteristics and realistic economic constraints called WAKE-NET. By coupling wind resource probability distributions with the Jensen wake model and costs associated with turbine, land, and cabling infrastructure, the WAKE-NET framework enables economical layout optimization. The results demonstrate that regional wind resource quality plays a decisive role in determining wind farm performance and profitability. Offshore sites, particularly Alaska, exhibit significantly higher AEP, capacity factor, and AEB compared to other regions. Onshore analysis further confirms that high wind regions such as Wyoming, Kansas, and Texas can achieve economically viable configurations, whereas low-wind regions may require policy support to remain financially feasible. Comparative analysis between the benchmark wake-ignorant and the wake-aware models highlights the necessity of incorporating wake effect during optimization. Neglecting wake interactions leads to overly optimistic energy estimates and sub optimal economic outcomes. Sensitivity analysis on the turbine count reveals the existence of an optimal farm capacity, beyond winch increased turbine density reduces capacity factor and profitability due to intensified wake losses and higher infrastructure costs. Finally, the investigation of multiple hub heights shows that while higher hub heights can access stronger winds, economic benefit depends on the balance between power generation gains and wake-induced losses. Overall, the findings emphasis that optimal turbine deployment at multiple hub heights is the key to achieving sustainable and profitable wind farm development.

The proposed WAKE-NET framework is designed to identify optimal combinations of wind turbine capacities, incorporate multiple hub heights to mitigate wake interactions, and quantify the minimum land area required to achieve efficient energy extraction. While the current framework captures key aerodynamic and economic drivers, practical wind farm planning involves additional factors that warrant further investigation. Further extensions of WAKE-NET will incorporate site-specific constraints such as high-resolution terrain topography, electrical network limitations, and grid congestion, enabling more realistic wind farm sizing and layout decisions. The framework can also be to incorporate turbulence-resolved wake models and fatigue aware economic optimization to evaluate the long term structural and operational impacts of wake-induced loading. Another extension that can be considered is the hub height dependent structural, transportation and foundation cost scaling to enable a more comprehensive techno-economic assessment of multi-capacity multi-hub height wind farm configuration. The current WAKE-NET framework assumes that all generated wind power is purchased by the microgrid immediately. In practice, however, power delivery is constrained by transmission capacity and network congestion. Integrating the WAKE-NET framework with a battery storage sizing optimizer would allow surplus energy to be stored during periods of limited grid availability, thereby improving grid flexibility and revenue potential. Additionally, the assumption of a fixed electricity price can be relaxed by incorporating dynamic pricing structures and accounting for inflation, thereby enabling more accurate long-term energy production and economic assessment.

## Acknowledgments

Research reported in this paper was in part supported by an Early-Career Research Fellowship from the Gulf Research Program of the National Academies of Sciences, Engineering, and Medicine. The content is solely the responsibility of the authors and does not necessarily represent the official views of the Gulf Research Program of the National Academies of Sciences, Engineering, and Medicine.